\documentclass[]{interact}

\usepackage[english]{babel}
\usepackage[utf8x]{inputenc}
\usepackage{apacite}

\usepackage{epstopdf}% To incorporate .eps illustrations using PDFLaTeX, etc.
\usepackage[caption=false]{subfig}% Support for small, `sub' figures and tables

\usepackage[longnamesfirst,sort]{natbib}% Citation support using natbib.sty
\bibpunct[, ]{(}{)}{;}{a}{,}{,}% Citation support using natbib.sty
\renewcommand\bibfont{\fontsize{10}{12}\selectfont}% To set the list of references in 10 point font using natbib.sty

\theoremstyle{plain}% Theorem-like structures provided by amsthm.sty

\theoremstyle{definition}

\theoremstyle{remark}

\usepackage{gensymb}
\usepackage{csquotes}
\usepackage{subfig}
\usepackage{xcolor}

\begin{document}
\newcommand{\jlc}[1]{{\color{blue}[JL: #1]}}
\newcommand{\rs}[1]{{\color{red}[RS: #1]}}

%\articletype{ARTICLE TEMPLATE}% Specify the article type or omit as appropriate

\title{Analyzing Behavior and User Experience in Online Museum Virtual Tours}

\author{
\name{Roman Shikhri\textsuperscript{a}\thanks{CONTACT R.~S. Author. Email: rshikhri@campus.haifa.ac.il} ,  Lev Poretski\textsuperscript{a},
Joel Lanir\textsuperscript{a}}
\affil{\textsuperscript{a}University of Haifa, Israel;}
}

\maketitle

\begin{abstract}
The disruption to tourism and travel caused by the COVID-related health crisis highlighted the potential of virtual tourism to provide a universally accessible way to engage in cultural experiences. 360-degree virtual tours, showing a realistic representation of the physical location, enable virtual tourists to experience cultural heritage sites and engage with their collections from the comfort and safety of their home. However, there is no clear standard for the design of such tours and the experience of visitors may vary widely from platform to platform. 
We first conducted a comprehensive analysis of 40 existing virtual tours, constructing a descriptive framework for understanding the key components and characteristics of virtual tours. Next, we conducted a remote usability study to gain deeper insights into the actual experiences and challenges faced by VT users. Our investigation revealed a significant disparity between users' mental models of virtual tours and the actual system behavior. We discuss these issues and provide concrete recommendations for the creation of better, user-friendly 360-degree virtual tours.

\end{abstract}

\begin{keywords}
360-degree Virtual Tours; Virtual Tours; Virtual Tourism; Cultural Heritage
\end{keywords}

% 1

\section{Introduction}
Virtual Tours (VTs) featuring 360-degree views of the environment have gained popularity in recent years allowing users to virtually explore cultural heritage sites such as museums and historical sites, as well as other real world environments such as real-estate properties and university campuses on their personal digital devices. A virtual tour is a realistic presentation of a physical location constructed by multiple 360-degree images or videos, usually shot with a 360-degree camera or stitched from a series of photographs \citep{bakre2017campus}.
The users can interactively navigate through the space, alter the viewing angle, and look around as if they were actually there.
360-degree VTs can serve as a way to view a location from afar and even serve as a particular representation of a place at a specific time in the past. 

The COVID-19 pandemic along with the lockdowns and restrictions it introduced which limited tourism activities, have made VTs an attractive option for individuals who want to learn about museums and cultural places remotely \citep{yang2021impact}. 
In conditions in which it is impossible to physically visit cultural locations and exhibitions, virtual tourism can partially recreate the experience by allowing the users to remotely view and interact with tourist destinations from the comfort and safety of one’s home \citep{adriyanto2015360}. Furthermore, VTs were found to increase the interest of the users to visit the actual locations of the tour  \citep{mohammad2009development} and to improve the learning outcomes and knowledge retention when they physically visit the location \citep{abidin2020students}.
Thus, 360-degree VTs provide an accessible solution for various types of potential users, from those who simply want to preview the location before they actually visit it, to others who may not be able to physically arrive to the site, either because they are far away, or because of other accessibility issues. 

%In addition to the convenience and cost-effectiveness of VTs compared to their physical counterparts \citep{Sylaiou_2008}, previous studies on virtual tourism have found that VTs can positively influence the psychological well-being of the users \citep{li2021study}, provide a sense of engagement, and reduce stress \citep{yang2021impact}. Moreover, VTs were found to increase the interest of the users to visit the actual locations of the tour  \citep{mohammad2009development} and to improve the learning outcomes and knowledge retention when they physically visit the location \citep{abidin2020students}.

Despite the growing popularity and usefulness of VTs, many VT visitors encounter usability issues that affect their user experience \citep{mohammad2009development, chiao2018examining, kabassi2019evaluating}. Furthermore, there is currently no standard way for a VT to be constructed, nor a standard way for the design of its components and user interface. Consequently, the attractiveness and experience for users using VTs vary depending on the particular design features and interaction mechanisms implemented in the specific tour. 
This lack of consistency and unified design knowledge negatively affects user experience and satisfaction \citep{mohammad2009development}. 

The overall goal of our work is to understand how to better design interactive virtual tours. As a first step to achieve this goal, this paper aims to gain a thorough understanding of the current practices and user experiences with VT applications. 
%in order to improve VT design, a thorough understanding of the current practices and user experiences with current applications is needed.
%Motivated by the existing gap in the knowledge on VT design, 
We first analyzed 40 existing VTs, evaluating them in the context of their features designed to facilitate the interaction between the system and the user. We focused on tours of museums and cultural heritage sites as a typical use-case for VTs. Based on this analysis, we created a descriptive framework to guide the creation and evaluation of VTs. Next, in the main part of this work, we conducted a remote user study to examine the behavior and experience of users browsing through two specific VTs. We used the think-aloud protocol to understand the way users perceive the interface and to unveil their misconceptions and pain points. 

Results offer an overview of how users use and perceive VTs.
We found that many participants struggled with their navigation, often experiencing a mismatch between their mental models of the system and actual system behavior. users did not always understand the way the system works and often expected the system to behave differently than it does. We also found that participants looked and used diegetic cues - information and elements that are embedded in the environment, as both navigation aids and information elements.  

The contributions of this paper are as follows:
\begin{enumerate}
    \item We provide a descriptive framework for virtual tours. The framework can be used to guide the design and evaluation of VTs
    \item We analyze user behavior and experience when using VTs, listing user pain points and usability issues.
    \item We provide concrete recommendations for the design of more usable VTs    
\end{enumerate}

The analysis and recommendations provided in this paper can benefit both industry practitioners and researchers to better understand and design VTs in order to provide enriching, pleasant, and useful experiences for virtual visitors. The paper is organized as follows: Section 2 described related work on virtual tours and navigation in virtual spaces. Section 3 describes the descriptive framework that stemmed from our analysis of existing VTs. Section 4 describes the methodology of our main study following its results which are listed in Section 5. Finally, section 6 discusses the results and lists the recommendation guidelines as well as the limitations of our study. Final conclusions are listed in Section 7. 

% contributions: 
% framework
%behavior analysis (pain points)
% design recommendations
%\rs{
%Our study explores the effects of different tour affordances on the user experience and reveals which factors influenced visitor behavior in virtual tours. A good user experience is a central aspect of the success of interactive products. The result will be improved user experiences in such tours and more potential clients. Thus, our main research question is: In 360-degree virtual tours, what are the main pain points users face and how is the experience of virtual visitors affected? . Our premise is that analysis of the user's behavior and experiences will help us to understand how different features and affordances in virtual tours affect behavior, pain points, usability, and user experience and how VTs can be designed with an improved user experience. Therefore, our research aims to provide the answers to the following research questions in order to fill in the gap:

%\begin{enumerate}
%\item 1. What are the current pain points of Virtual Tours?
%\item 2. What are the key design guidelines a 360-degree Virtual Tour needs to address?
%\end{enumerate}
%}

\section{Related Work}

\subsection{Virtual Tours}
Virtual tourism presents users with a realistic experience of a  location or destination through the use of technology \citep{cho2002searching}. One of the primary forms of virtual tourism is a 360-degree virtual tour. Lerario and Maiellaro \citep{lerario2020virtual} defined VTs as representations of the natural environment in virtually recreated pseudo-3D spaces consisting of an interconnected series of 360-degree images that map and reproduce an actual physical place. In HCI research, these methods of capturing the environment are sometimes defined as a “scanned reality” \citep{Ipsita2021-dg}. 
In image-based VTs, the users are usually placed at the center of the 360-degree image. They can change their angle of view by dragging a mouse or pressing the directional keyboard keys \citep{lerario2020virtual}, similarly to how changing perspectives is typically implemented in conventional desktop-based virtual spaces. Additionally, the users can move between the images, with  smooth transitioning creating an illusion of movement through space. Google Street View \footnote{https://www.google.com/streetview/} is probably the most widely known application of such technology.

\par A substantial body of knowledge is developed around designing and evaluating VTs, from technical implementation approaches \citep{napolitano2017virtual, Lluch2005-ih}, to user interface design \citep{lerario2020virtual}, to exploring navigation modalities \citep{mohammad2009development}. Furthermore, the ongoing development of immersive digital technologies leads to the development of new and creative approaches for VT design. For example, omni-directional videos can be used instead of static images to support the VT experience (e.g., \citep{Sugimoto2020-rh,argyriou2020design}). Additionally, 
%physical props may be combined or connected to their virtual representations for a more engaging experience \citep{Lopez2021-fi}. 
several studies examine virtual reality as a way to access and interact with VTs \citep{giangreco2019virtue, fineschi20153d, kersten2017development}. However, due to the universal accessibility of Web technologies, overwhelmingly, VTs continue to use 360-degree images stitched together, and the predominant way users access VTs is through the Web. Thus, in the current paper, we focus on image-based tours that are accessed through a standard Web-based interface. 

%VTs have steadily increased in popularity in recent years, which is further sped up by the ongoing health crisis. 

\par The importance of designing for engagement and immersion  was found to be critical for a positive user experience \citep{mohammad2009development, namsong2009virtual}. In addition, VTs should implement a range of functionalities, such as interactive maps, floor plans, and external indications of locations, which allow the users to engage in planning their route and defining personal points of interest \citep{lerario2020virtual}. Such aids help to increase the sense of continuity of the environment and the feeling of increased control over the experience \citep{Scuri2020-ug}. The information supplementing the user experience, such as texts and multimedia elements that describe the artifacts of the toured place, were also found to be an important component of the tour design \citep{koehl2013documentation, Thom-Santelli2010-ka}. 
Finally, some works have explored the narrative component in the context of virtual experiences in general \citep{Benford2009-mu}, and  VTs in particular \citep{Damiano2013-vf}. In order to have a more complete experience, the users can be guided through the tour according to a pre-designed user journey, informed by a coherent narrative  \citep{Benford2009-mu}.

\par While a substantial body of knowledge exists on the design and evaluation of novel VTs, not much is known on the current experiences of regular users beyond the experimental solutions. Moreover, the existing recommendations are fragmented and distributed between different domains of study, such as HCI, Cultural Heritage, Digital Humanities, and Multimedia research. Our work aims to empirically investigate user-driven design aspects of virtual tours for human-computer interaction.  

\subsection{Navigation in virtual spaces}
\par One of the key factors related to our work is the importance of effective navigation and movement in VTs. 
%This includes figuring out the optimal speed of transition between the tour scenes and providing effective means of orientation in space \citep{mohammad2009development}. In addition, auxiliary visual aids, such as waypoints and directional arrows overlaid on top of the tour’s user interface, can significantly increase navigation efficiency \citep{Elmqvist2008-we}.
Navigation in virtual environments, whether desktop-based or immersive virtual reality environments, is a well-explored topic in HCI literature. For our purposes, navigation can be conceptualized as wayfinding, which refers to the condition in which navigation is not the primary goal of the interaction but is essential to achieve some other specific task \citep{Darken1996-qu}. Effective wayfinding is dependent on the ability of the users to create and maintain cognitive maps of the virtual spaces \citep{Boff1986-vd, Tolman1948-yg}. Cognitive maps are a part of the overall human spatial cognition and describe the mental representation of the environment \citep{Tolman1948-yg}. They consist of a set of landmarks and their spatial relations and are heavily used in navigating both physical and virtual environments.  
However, the human ability to form cognitive maps of virtual environments is often not as good as in a physical settings due to several factors generally related to the reduced fidelity of the virtual spaces \citep{Elmqvist2008-we}. These factors include reduced visual fidelity of an environment, unnatural movement and navigation controls, and lack of additional sensorial stimuli, such as sense of balance, temperature, and touch \citep{Chittaro2004-fy}. Consequently, users often experience disorientation and increased cognitive load during navigation in virtual environments, warranting the implementation of additional features and characteristics that aid navigation \citep{Elmqvist2008-we}. 

\par Elmqvist et al. discuss navigation with motion constraints and suggests three main principles for such navigation to be effective \citep{Elmqvist2008-we}. First, all the important points of interest in the environment need to be presented and reachable by the user. Second, continuity of motion is important in order to promote adequate understanding of the environment (see also \citep{Malekmakan2020-jm}). Finally, it is important to enable users’ control of movements, providing local deviations and implementing self-motion cues for the users to promote spatial learning (see \citep{Lim2020-ic}). Pertaining to 360-image based VTs, the main challenge is in the implementation of continuity of motion. Due to the way in which VTs are engineered, the standard methods of continuous movement in virtual environments, such as smooth movement while a keyboard key is pressed, are impossible to implement. Thus, continuity of motion in VTs is typically achieved through implementing various transition effects between the images when the user initiates movement from one image to another, creating an illusion of smooth movement. 

%\par A substantial body of work has been devoted to designing and evaluating various navigation techniques and metaphors that can best support effective wayfinding and navigation. Bowman and colleagues have developed a taxonomy of first-person-based navigation and movement methods in virtual environments \citep{Bowman1997-aj}. Applying Bowman’s et al., taxonomy on VTs, we describe three aspects of navigation in VTs: (1) The method by which the user selects the goal position of the movement (\textit{Direction and target selection}). In the case of VTs, this is done using a discrete selection of direct targets embedded into the virtual world (e.g., points of movement at the center of each 360 image). (2) The way velocity and acceleration is determined (\textit{Velocity and acceleration selections}). In the case of VTs, automatic adaptive velocity is employed depending on the distance between the user and the target of movement. (3) \textit{Input conditions} refer to ways in which the user or system specifies the beginning time, duration, and end time of the travel motion. In VTs, every movement is discrete with the start movement input being user driven, and the system controlling the duration and end of each movement. 

\par 360-degree VTs use discrete movement navigation from one point to another as a basic movement metaphor. For such discrete movements, automatic adaptation of velocity to distance-of-travel based on nonlinear speed is a well-known design requirement \citep{Ware1988-tg,Bowman1997-aj}. In practice, this usually means that the larger the distance between the user and the intended point of movement, the faster the transition is to this point. Several ways to implement the non-linearity of the motion were proposed (e.g., \citep{Mackinlay1990-ge, Song1993-bl, Tan2001-nt}). Non-linearity of movement is an inherent property of many 360 VTs. For example, in the VTs created using Google's technology, the time of travel takes an equal amount of time regardless of whether the user moves to a close or far location. 
%Subsequently, high-level navigation and locomotion design principles result in several more specific wayfinding techniques, such as \emph{flying}, \emph{scene in hand}, \emph{eyeball in hand} \citep{Ware1988-tg, Ware1990-xo}, \emph{raycasting} \citep{Hinckley1994-oa}, and \emph{world-in-miniature} \citep{Elmqvist2007-rs, Truman2020-ve}. 

\par The effectiveness of navigation and wayfinding techniques, to a large part, depends on how designers implement additional navigational aids. Two types of navigational aids are identified: \emph{guided navigation} and \emph{motion constraints} \citep{Elmqvist2008-we}. Guided navigation refers to various directional signs, directional arrows, and interactive maps (e.g., \citep{Chittaro2004-fy}). 
%Another way to implement guided navigation is to create historic paths that are overlayed on the environment and represent previous movements of the users \citep{Ruddle2005-bs}. Finally, embodied virtual guides also can help navigation and wayfinding by actively directing the users on where to go (e.g., in \citep{Chittaro2003-oc, Yoon2005-gs}).
%JL - this can be moved somewhere else
%Navigation assistance of the virtual guides can be reference-based, direction-based, or turn-by-turn. The research has shown that reference-based navigation leads to a more effective formation of cognitive maps \citep{Kuo2020-iw, Vinson1999-vg, Singh2021-ke}.
Motion constraints 
%are a capable alternative (or complement) to guided navigation in design for user wayfinding \citep{Elmqvist2008-we}. It 
refers to the different ways to constrain users’ movements such as 
%While different degrees of motion constraint can be implemented, existing works generally distinguish between  fixed-gaze and motion-weighted gaze techniques \citep{Wernert1999-pr}. Fixed gaze refers to an 
constraining the movement along a pre-determined path with or without having the freedom of gaze direction \citep{Galyean1995-eo,Wernert1999-pr}.
%Motion-weighted movement refers to having freedom in gaze direction \citep{Galyean1995-eo, Wernert1999-pr}. Motion-weighted movement can either automatically adjust users’ gaze direction toward a point of interest when the movement is slowing down or even change the direction of the gaze depending on the current direction of movement \citep{Elmqvist2007-rs, Wernert1999-pr}. Instead of restricting the users' movement completely, 
Motion constraints can also be implemented in the form of invisible “walls” or other obstacles that prevent the users from steering from the desired course \citep{Hanson1997-dg}. 
%In such a way, the users retain the freedom of movement within the environment, while the designers still have a mean to guide the users’ movement into the desired direction. 
\par While existing research on navigation in 3D environments have extensively examined possible design solutions for navigation and wayfinding in virtual environments, it is currently unclear how the proposed techniques translate to the “scanned” environment of 360-degree VTs. Due to the significant differences in construction and presentation of such environments, leading to the somewhat unique way of implementing motion (or, rather, transition) within 360-degree  VTs, it is unknown whether existing VT solutions lead to a good user experience. Our study aims to contribute to the current body of knowledge on user experience of navigation in 360-degree VTs. 

%3

\section{Descriptive Framework for 360-degree Virtual Tours}
% ^{\circ}$ = degree
In order to better understand the design structure and characteristics of VTs, we reviewed 40 existing VTs of museums and cultural heritage sites, examining their features, cues and affordances that facilitate interactions between the users and the system. 
To decide which VTs to review, We first constructed a large list of VTs by going over virtual tourism portals and by searching the Web for the terms: "virtual museum tours". We then removed outdoor and other types of less relevant tours leaving only classical museum VTs. Forty tours were then chosen of relatively known museums that vary across multiple platforms, interaction modalities, and informational content. 

We analyzed each of the tours according to its various features, visual cues and interaction design.  We started by analyzing 10 tours using a bottom-up approach, forming initial categories of features and cues (e.g., does the tour have a map? what kind of map does it have?). We then analyzed all tours according to these categories, listing all features and cues in a large spreadsheet, slightly changing and adding categories when we saw they needed readjustment. Finally, we went over the various features in the spreadsheet, finding commonalities and differences.
Following this review, we identified several common patterns in the interaction design of VTs, creating a high-level framework of design dimensions that can be used to guide the creation and evaluation of VTs. 

Common to all VTs is that they rely on what we term  \textit{Spatial Navigation Points} (SNP). SNPs are interface navigation points that serve as movement points for the users. They are usually presented as a visual icon or clickable point in the environment. When users click on a specific SNP, the system transfers them into this specific location, displaying a 360-degree image centered in this point. SNPs prompt users to move by clicking them in order to advance in the room, to explore the room, or to examine the nearby exhibit. SNPs are usually determined according to the mapping of the space: each SNP is mapped to a spatial position that the tour designer decided to take a 360 image from. SNPs are sometimes referred to as \textit{points of interest} (POIs). However, we define POIs as locations that someone may find useful or interesting. Thus, there might be POIs with no SNP defined near them, and there might be SNPs that are not POIs (e.g., SNPs in the middle of a room, or ones that lead the user through a corridor).

We identified three main conceptual dimensions for VTs: \emph{navigation, information presentation, and proactiveness}. Table \ref{tab:framework} summarizes the framework's main elements.  
\begin{table}
  \caption{Descriptive framework of virtual tours}
  \label{tab:framework}
  \begin{tabular}{p{0.15\linewidth}p{0.25\linewidth}p{0.5\linewidth} }
    \toprule
    Dimension&Sub-elements&Description\\
    \midrule
    Navigation & Freedom of Movement& How many options and places the users can choose to move to within the tour's environment.\\
     & Movement Transition& The fluidity of the transition when moving around the space.\\
    & Spatial Orientation & The design aspects of the virtual tours that support one's sense of direction and location within the environment.\\
     & User Controls & The way in which the users' input mechanisms and interaction are implemented. \\ \hline
     Information Presentation & Visual Information Cues & The visual elements that provide additional information, options, and interaction opportunities.\\
     &Multimedia support&Additional audio-visual elements\\ \hline
     Proactiveness&Navigation proactiveness&The amount of proactiveness of the system in aiding the navigation\\
     &Interaction proactiveness&The amount of proactiveness of the system in providing information to the user\\ 
     %\hline
     %Embedded interaction&&Interactive elements that are embedded within the tour's environment.\\
  \bottomrule
\end{tabular}
\end{table}
%\textbf{}

%1 ----------------------------------------- Navigation ------------------------------------------------------------
The \textbf{navigation} dimension describes the affordances built into a VT that support the user's ability to navigate and move around the environment. We identified the following aspects of VT navigation:

\begin{itemize}
   \item \emph{Freedom of Movement}. This aspect refers to the extent to which the users can move within the tour's environment. Freedom of movement depends on the number and placement of SNPs, and the scale of the tour's environment. That is, the ratio of SNPs to the size of the environment will determine the ability of the users to explore and navigate the space. 
   %Consequently, the lowest freedom of movement is associated with  tours in which only a single SNP exists for each location, essentially making it a "no movement" tour. Such design can work with relatively small exhibitions comprised of only a few small rooms. 
   %More SNPs lead to more continuous movements and better transitions between the points, allowing users to reach more locations and areas.
  
   \item \emph{Movement Transition}. Movement transition refers to the fluidity of the movement around the space, ranging from a discreet appearance of the chosen SNP on the one end, to a gradual and smooth transition between the points on the other end. Movement continuity is known to be important for the spatial understanding of the environment \citep{Elmqvist2008-we,Malekmakan2020-jm}. The movement transition is affected by the overall distance between the movement points, and the design of the transition effect. In addition, the quality of the image stitching can also affect the resulted perception of movement transitions. 
   
  \item \emph{Spatial Orientation}. Spatial orientation refers to the design aspects of VTs that support users' overall orientation within the environment. One of the common ways in which the spatial orientation of the user can be supported is the inclusion of a persistent ego-centric mini-map, in which the user can see his or her updated location in the schematically represented environment. Other spatial orientation options that are commonly used in VTs include a 360-degree panoramic overview of the environment, aerial top-down view of the points of interest, simple lists of available places, and visual showcases of key or popular points of interest using snapshot images. 

  \item \emph{User Controls}. User controls describe how users' input and interaction controls are implemented in a VT to support navigation. The input controls may involve physical input devices such as a keyboard, mouse, or a joystick that map input commands to navigation actions (e.g., mapping the arrow buttons in a keyboard to movement in the VT). An interaction control commonly used are "soft" buttons presented on screen, which can be clicked with a mouse or touched if the tour is viewed on a touchscreen device. Similarly, dragging the mouse or finger can be used for rotation, allowing to change the users' gaze angle. 
  %Another common navigation controls are the visual markings of the SNPs, serving as interactive anchors for transition to the point. 
  
\end{itemize}
% --------------------------------------------------------------------------------------------------------------------------

%2 ----------------------------------------- information presentation ------------------------------------------------------------
The second dimension in the framework is \textbf{Information Presentation} which refers to the means in which the system enriches or augments the reconstructed environment of the tour with additional layers of information. 
%We distinguish between two information presentation types:
\begin{itemize}
  \item \emph{Visual Information Cues}. These are visual elements laid out on top of the environment that help users understand their options and interaction opportunities related to real items. For example, a red icon overlaid on top of the exhibit may indicate that the user may access additional information about this exhibit by clicking on the icon. Visual cues can vary in color, size, and form and can be implemented in the form of clickable labels, icons, dialog windows, or other interactive media elements.   

  \item \emph{Media Content}. These are additional media information about the museum exhibits. VTs can include a variety of media types and contents including text, images, video, 3D models and replicas, and others. 
  %These media contents can be accessed through visual information cues, or via other means (e.g., a textual description can be overlayed on top of the environment).
\end{itemize}
% --------------------------------------------------------------------------------------------------------------------------

%3 ----------------------------------------- proactiveness ------------------------------------------------------------
The third dimension in the framework is \textbf{Proactiveness}. This concept relates to the degree to which the system tries to aid the user to perform or follow desired actions by predicting the intentions of the user. A completely proactive system will take all the decisions and actions from the hands of the users, similar to how a tourist guide may manage a group in a tightly controlled manner, designing fully guided experiences through the cultural location. Conversely, a system that lacks proactiveness will provide the user with full control, similar to how visitors may freely walk and browse museum environments.
We identified two types of proactiveness in VTs:
\begin{itemize}
  \item \emph{Navigation proactiveness}. This concept refers to a systems that automatically transfers users across different points of interest, instead of waiting for explicit navigation input from users. Such systems are trying to predict the users' desired route and execute the appropriate transitions between different locations. 
  
  \item \emph{Interaction proactiveness}. This refers to the system automatically triggering the interactions between the users and different components of the VT - such as information containers, interactive UI elements, and system dialogues depending on the contextual appropriateness of such actions (e.g., see \cite{lanir2011examining}. For example, the system may automatically present a window with a textual description of the exhibit when the user gets close to it. 
\end{itemize}

% --------------------------------------------------------------------------------------------------------------------------

%4 ----------------------------------------- interactivity ------------------------------------------------------------
%The final dimension in the framework is \textbf{Embedded Interactivity} which describes the extent to which the users can directly interact and manipulate the environment and elements within it. For example, the users may be able to interact with the exhibits within the environment by highlighting and contouring them, measuring their proportions, rotating or otherwise manipulating them, etc. 

%4

\section{Methodology}
%what did we do... in two sentences
We conducted a remote qualitative study to examine users' behaviors and attitudes when using 360-degree virtual tours. The study involved the participants visiting a VT under the researcher's guidance while following the think-aloud protocol. In the current section, we describe the research methodology in detail. 

\subsection{Participants}
Fourteen participants took part in our experiment. We wanted participants who are interested in cultural heritage and museums since these are the expected users of these kind of VTs. Thus, eight of the participants were travel and cultural heritage hobbyists recruited from a travel guidance course given at our university. In addition, six other participants indicated a general interest in cultural heritage sites, and were referred to by the other participants. The average age of the participants was 53.7 years (range  28 to 71). Five of the participants identified themselves as female and nine as males. The participants have come from a diverse educational and vocational background, from elementary teacher, to data scientist, to former vice president of a technology company. The average reported technological proficiency of the participants was 3.85 on the Likert scale of 1-5 (SD=0.95). Participants have indicated moderate interest in experiencing VTs, with an average of 3.07, on the scale of 1-5 (SD=0.92). Their interest in tourism was 4.71, on a scale of 1-5 (SD=0.83). Reported familiarity with VTs showed an average of 2.28 (SD=1.20), with 9 out of 14 participants having tried them in the past.

% Only XX have indicated that they have experienced VTs more than once. 
% \lp{Roman, can you please fill in the details that are currently indicated as XX? Also, you can add the rest of hte demographic data from your Google Form}

\subsection{Procedure}
\par We asked the participants to use their home personal computers equipped with a mouse. At the start of our experiment, we contacted our participants remotely via the Zoom video-conferencing software. After the initial introduction, we explained the goals of the study and the research protocol, including the procedures related to our recording and handling of the personal data collected during the experiment. The participants were then invited to sign an online consent form and fill in a demographic questionnaire via a shared link. We then asked the participants to share their screen with us and initiated an audio-visual recording. 

\par Upon initiating the recording, we sent the participants a link of the VT to be used. The order of the tours was counterbalanced. The participants were requested to enter the tour and follow the guidelines of the researcher following a standard think-aloud protocol. During the tour, the participants were encouraged to experience the museum as they see fit, freely looking around, engaging with every museum artifact they were interested in. Participants were encouraged to communicate their actions and thoughts and say out loud what they think about the interaction in the VT. The participants were requested to operate every aspect of the VT system using only the mouse. The participants were free to explore the tours for an unlimited amount of time. The average time participants spent in a tour was 31 minutes.

\par After participants verbally indicated that they had have experienced the tour to their full satisfaction, they were asked to fill out the post-experimental questionnaire capturing their user experience and level of satisfaction with the tour. Finally, we interviewed the participants, allowing them to express their general opinions and share their insights on various aspects of the VT experience. 

\subsection{Virtual Tours Chosen for The Study}

\par Finding the tours to use in our experiment was a challenging task. Using the list of 40 tours assembled at the initial stage, we iteratively reviewed existing platforms and specific tours looking at their particular features and interaction mechanisms. We ended up choosing two tours of different platforms. 
%\rs{Our criteria for choosing the two platforms was universality and feature-set variety. As for choosing the virtual tours, our criteria was navigation option diversity and museum commonality.}
The first platform, Google arts and culture platform, was chosen because of its ubiquity, as many existing VTs today use the Google's platform and its design and navigational affordances are often copied by other VTs. The second platform, Matterport, was chosen as a new leading platform that creates engaging, yet standard VTs. The two tours we chose were more or less representative in their spatial coverage, the number of SNPs and the number of information cues provided about the exhibits: 

\begin{enumerate}
\item \textbf{Google arts and culture platform} - Bode-Museum, Staatliche Museen zu Berlin, Germany \footnote{https://artsandculture.google.com/partner/bode-museum-staatliche-museen-zu-berlin}. 
\item \textbf{Matterport platform} - European Museum of Modern Art, exhibition of works by sculptor Grzegorz Gwiazda on display at the European Museum of Modern Art, housed in Gomis Palace in Barcelona, Spain \footnote{https://matterport.com/gallery/european-museum-modern-art}.
\end{enumerate}

\par
\begin{figure}%
    \centering
    \subfloat[\centering Matterport Platform ]{{\includegraphics[width=4.5cm,height=3.18cm]{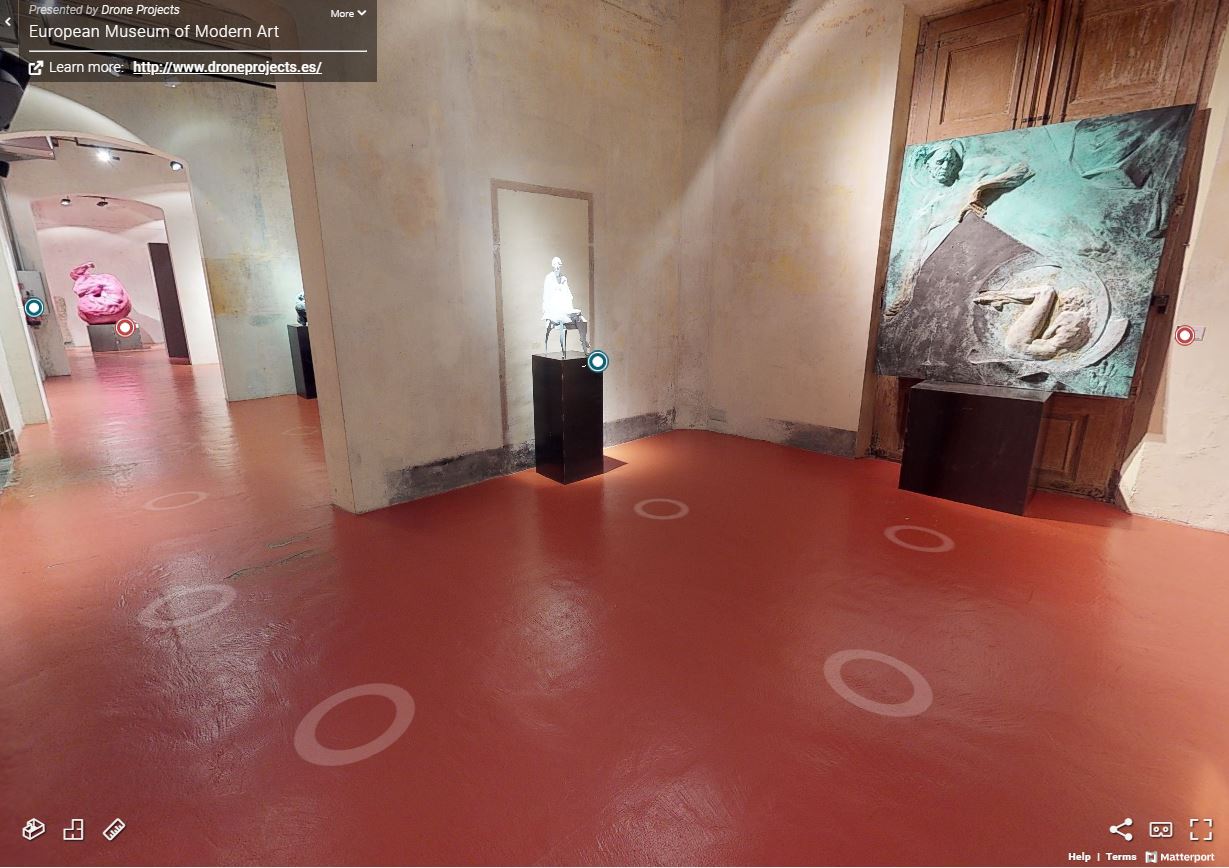} }}%
    \qquad
    \subfloat[\centering Google Arts and Culture]{{\includegraphics[width=4.5cm,height=3.18cm]{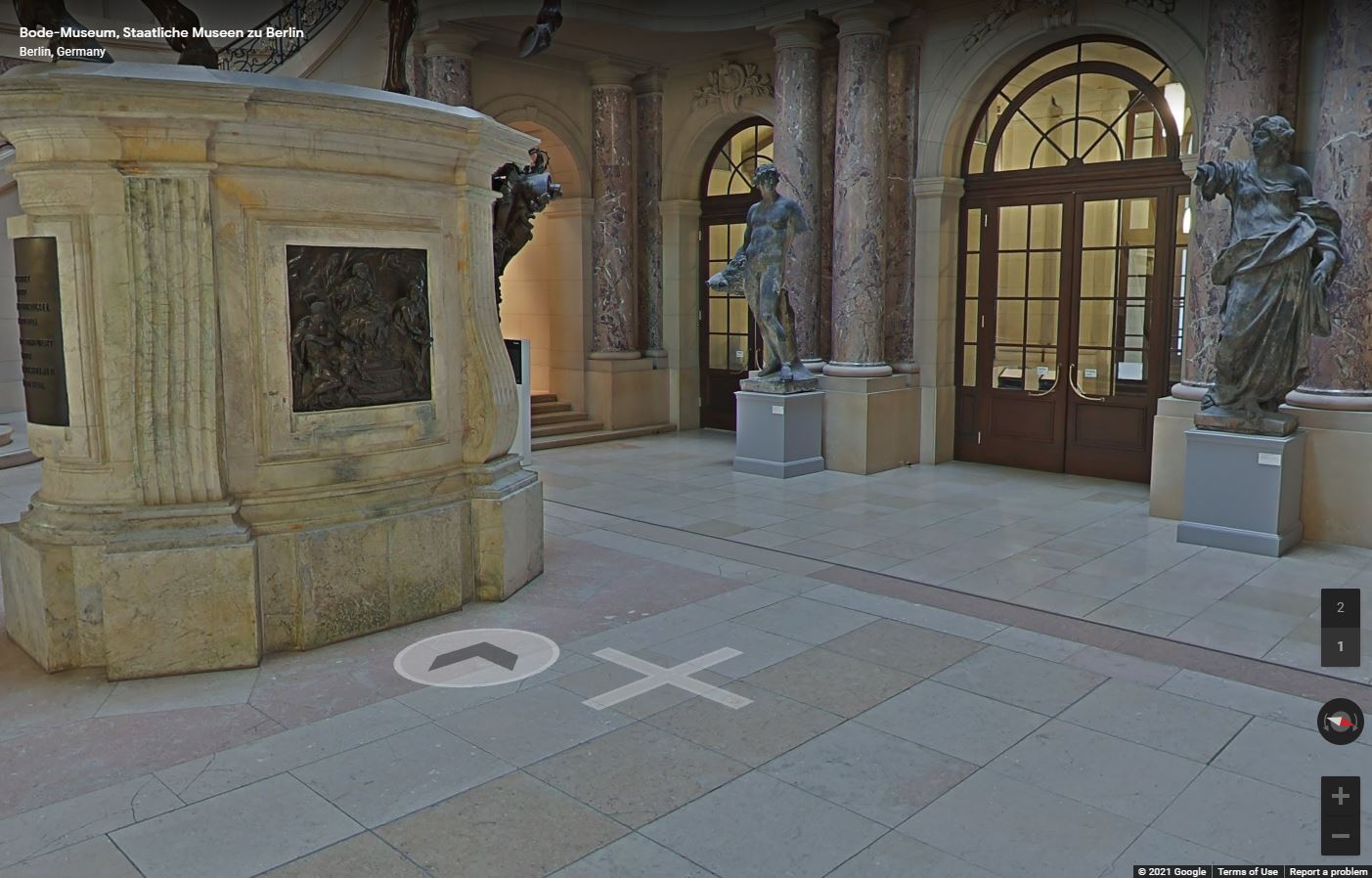} }}%
    \caption{Example of the two platforms used in the study}%
    \label{fig:snp-comparison}%
\end{figure}

 Example screenshots of both VTs are presented in Figure \ref{fig:snp-comparison}. Both tours base their navigation on SNPs, meaning that the users can only go from one SNP to another. In both tours, if users press on any point in the screen. the system moves the user to the SNP closest to the mouse press. The main difference between the tours is that in Matterport, all available SNPs are shown, while in Google, a directional arrow is presented when the mouse hovers in the bottom part of the screen, with the relevant SNP appearing when the mouse cursor is within its vicinity. In addition, the Matterport platform has a layout map of the museum that can be accessed using a button on the bottom left part of the screen.

%affordances of the settings : two types of arrows in google with images, matterport affordances too

\subsection{Data analysis}

\par We used a bottom-up approach to the qualitative analysis, following the guidelines for thematic analysis adopted for working with visual material by Suchman and Trigg \citep{Suchman1992-yq}, and aiming to provide a rich characterization of participants' interactions with 360 VTs. We watched all video recordings, identifying the key events and incidents around the participants' interactions with the VTs. Primarily, we focused our effort on analyzing the pain points and struggles of the participants, and identifying any ambiguities in the design of interaction that negatively affected users' experiences. To aid our understanding at this point, we created rich notes that captured the key aspects of the experimental process, accompanied by the screenshots and audio excerpts from the videos. 

\par In addition to the direct observations of the participants' behavior, we transcribed the think-aloud content (participants' verbal comments during the interaction) and interviews with the participants to further aid our analysis. These materials along with the notes have subsequently served to help inform the initial coding categories around the key phenomena of interest for the subsequent analysis. 

\par At the next stage of the analysis we iteratively refined the coding scheme, organizing our codes into related categories and noticing the recurring patterns. We have engaged in weekly discussions to review any emerged uncertainties and irregularities and to come to an agreement regarding our interpretation of participants' behaviors. We deemed our analysis complete after each new video stopped providing novel insights that could further enrich our understanding. In the following section, we provide a detailed description of our findings. 

%5

% 5    RESULTS %%%%%%%%%%%%%%%%%%%%%%%%%%%%%%%%%%
\section{Findings}
The results of our study offer a detailed overview of how users perceive and interact with virtual tours, identifying and describing the main pain points that our participants encountered throughout the tour visit. 
During the analysis, we examined the data gathered and paid particular attention to various aspects such as common behaviors, actions and their causes, and the context in which these actions occur. We documented the participants' reasoning and reflections describing their VT experiences, providing illustrative quotes for each important aspect of the user experience. We pay particular attention to functionality and the context of use that was frustrating or confusing to the users as well as enjoyable aspects of the VT experience. We frame our findings according to the categories in the framework presented in Section 3, focusing on the first two dimensions: navigation and information presentation.
%We also extracted and dived into things we had never heard or seen that users might do and contradictory feedback and results. 

\subsection{Navigation}
As described in Section 3, a core part of any VT is the users' ability to freely navigate through the environment. In this section, we focus on the various aspects related to the navigation experience of the participants during the study. 

\subsubsection{Navigation Strategies} 
\hfill\\
% navigation method 1 - clicking on exhibit/close to the exhibit/between points/point on the screen. not clicking on SNP ! 
During the study we observed how users employed two different methods of navigation. In the first method, which we term \emph{exhibit-based navigation}, users clicked on a specific point on the screen in order to move there. Mostly, such points were associated with specific exhibits and other objects in the environment.
%hence our definition as \emph{exhibit-based}. 
For example, P9 explained clicking on a painting: \textit{ "I see a painting I like from afar and I want to move straight to it"} (P9, 67 y.o.). In both platforms examined in the study, after registering a click, the system transfers the user to the closest SNP in the direction of the specified clicked location. However, this does not necessarily bring the user directly to the exhibit as there may be several SNPs on the way. This created inconsistencies between how  participants expected the system to act and the actual system behavior. 
%For example, when the users clicked on a statue, a painting, or any other item, they generally expected to appear directly in front of the item. In reality, the system could not support such high navigation fidelity. 
As a result, participants often felt confused and lost in the environment, 
as one of the participants commented: \textit{"I clicked on the statue and it moved me to a different place, I can’t understand why it happens"} (P5, 66 y.o.). 

% navigation method 2 - clicking on the SNP only ! 
In the second method of navigation, which we term \emph{point-based navigation}, participants navigated by directly clicking on the SNPs on the floor. In this case, the system reacted accordingly, moving users to the appropriate SNP. Such way of navigation reflects close mapping between system- and user- navigation models, allowing the user to go directly to the location she  intended. However, this navigation method seemed to be less intuitive for most participants, with less participants navigating using this strategy, at least at start. The downside of point-based navigation is that users are forced to search for available SNPs rather than paying attention to the tour itself. This often caused participants to become frustrated when they could not find a SNP in the direction they wanted to go, or an SNP that enables them to move to a desired exhibit.

% >> how the two navigation methods was reflected among users
Most users started their session using exhibit-based navigation naturally assuming that the system will react in an appropriate way. However, users quickly realized the inaccuracies of this type of navigation, leading to a higher use of point-based navigation. \textit{"I was confused at first. Now I realize I can only get to the marks on the floor"} (P12, 33 y.o.). In many cases, however, users  alternated between the two methods.

%\rs{Only 50\%  (7 out of 14) of the users replied positively to a question "Would you say the museum's navigation level is satisfactory to you?   }

% 5.1.1  END OF Navigation Strategies. %%%%%%%%%%%%%%%%
%%%%%%%%%%%%%%%%%%%%%%%%%%%%%%%%%%%%%%%%%%%%%%%%%%%%%%%
%%%%%%%%%%%%%%%%%%%%%%%%%%%%%%%%%%%%%%%%%%%%%%%%%%%%%%%

%%%%%%%%%%%%%%%%%%%%%% 2 %%%%%%%%%%%%%%%%%%%%%%%%%%%%%%
%%%%%%%%%%%%%%%%%%%%%%%%%%%%%%%%%%%%%%%%%%%%%%%%%%%%%%%
% 5.1.2    Difficulties in navigating through space %%%
\subsubsection{Difficulties in navigation and spatial orientation}
\hfill\\  
The use of the exhibit-based navigation strategy, coupled with the system's inability to fully support it, often led to user disorientation in space. Participants did not understand why their actual movement was misaligned with where they clicked and the system provided no indication or feedback on its navigation behavior.

%For example, one of the participants expected that by clicking on the statue she will come next to it. However, because the system actually moved the participant to the SNP that it calculated as most close to the location of the mouse click, the participant found herself in a different place unexpectedly (P3 (32 y.o., female)). 

\par
\begin{figure}[h]
      \centering
      \includegraphics[width=0.35\linewidth]{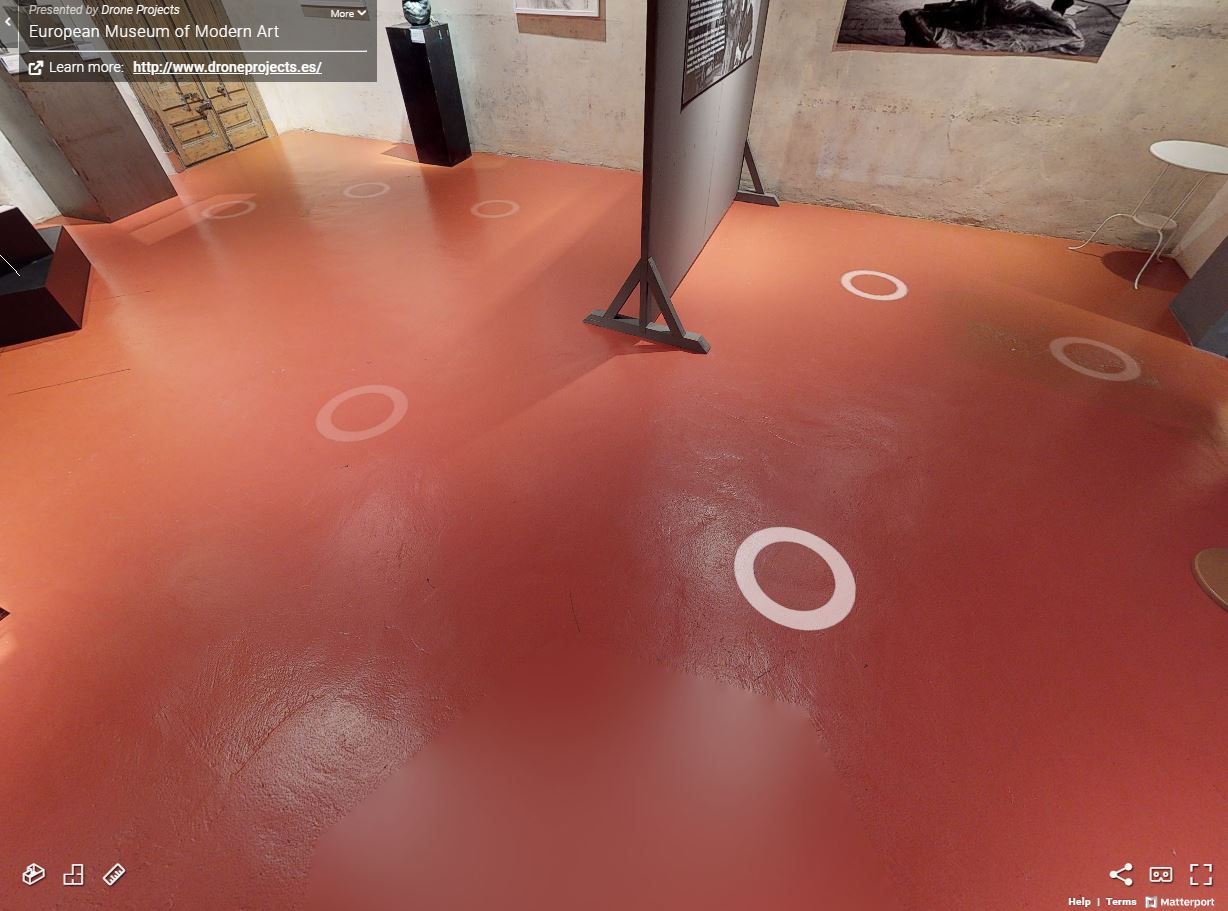}
      \caption{Users looking down on the floor in order to find the next available SNPs.}
      %\Description{SNP search results in excessive focus on the floor}
      \label{fig:focus-stuck-on-floor}
    \end{figure}

 While point-based navigation is more aligned with the way the system is designed, it also created some difficulties. In particular, we observed that point-based navigation caused participants to change their virtual gaze direction and look at the floor, searching for available SNPs (see Figure \ref{fig:focus-stuck-on-floor}). This happened with 13 of the 14 participants. Often participants remained in this gaze direction for a long period of time. This had a negative effect on their overall spatial awareness since they viewed a more narrow part of the scene. 
For example, P13 reported that he did not realize how to change his gaze to the normal point of view, complaining: \textit{"I look at the floor too much, now I see everything from high above. I lost my sense of navigation in terms of where I am" (P13, 71 y.o)}.

%Joel: this paragraph is about the visibility of SNPs. currently removed, but can be added somewhere else

%Using point-based navigation was more difficult in the Google platform in which not all SNPs are visible at once. In fact, in the Google's system, users can only see one navigation point at a time - the one that is closest to the mouse pointer's current position (see Figure \ref{fig:snp-comparison} left). This implementation of SNP's presentation considerably increased the time required from the users to realize that navigation is performed using SNPs rather than the exhibits, and caused participants to spend a lot of time and effort simply searching for SNPs with their mouse cursor. On the other hand, in the Matterport platform, all SNPs were visible at once (Figure \ref{fig:snp-comparison}, right). Consequently, when using Matterport,  users quickly understood that SNPs serve as the anchors for their movement in space. Moreover, in the Matterport system, most users felt more confident about where and how they could move, and devoted more time to the museum experience itself.   

\begin{figure}[!htb]
    \centering
    \subfloat[\centering Start position - a participant  pressing the left statue expects to view it. ]{{\includegraphics[width=0.35\linewidth]{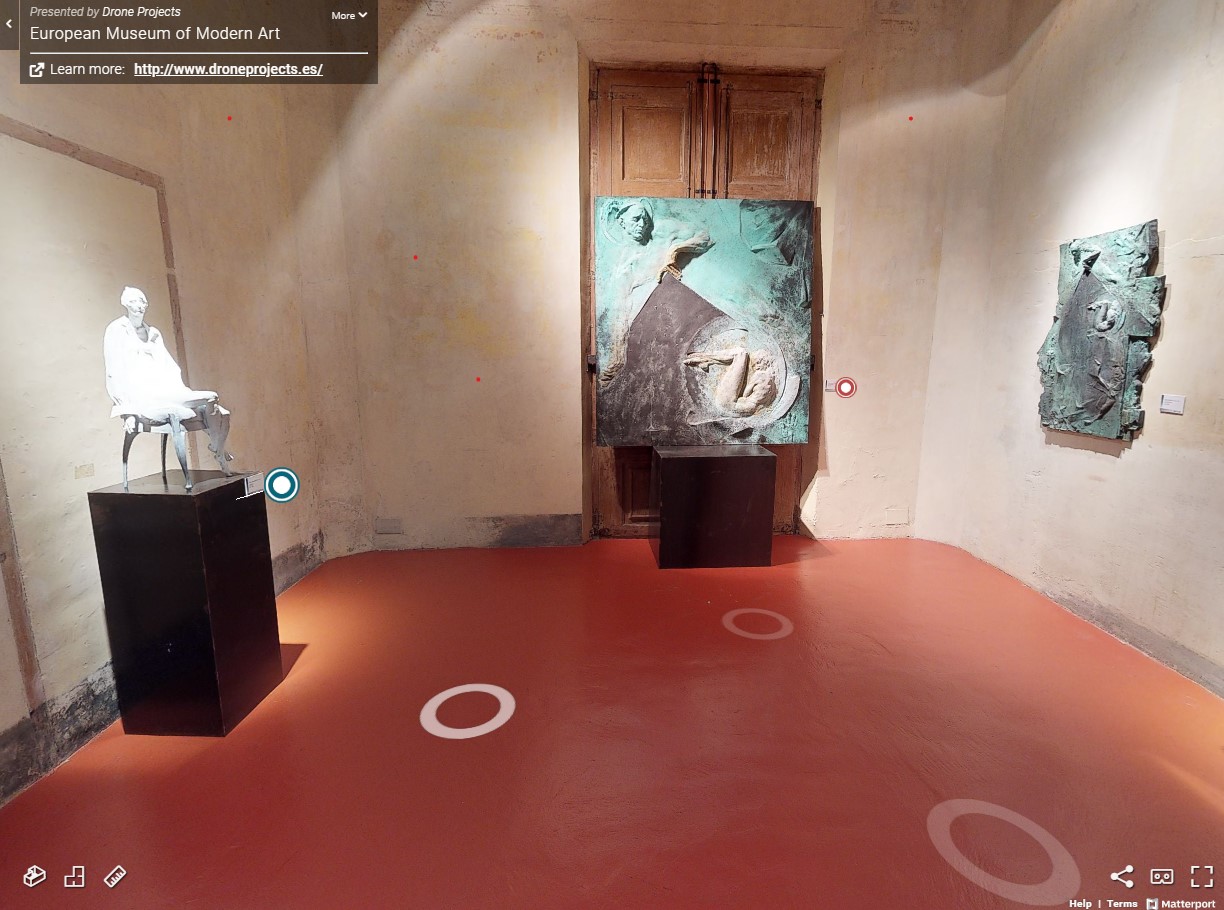} }})%
    \qquad
    \subfloat[\centering End position - after pressing the statue the participant cannot see it, creating disorientation.  ]{{\includegraphics[width=0.35\linewidth]{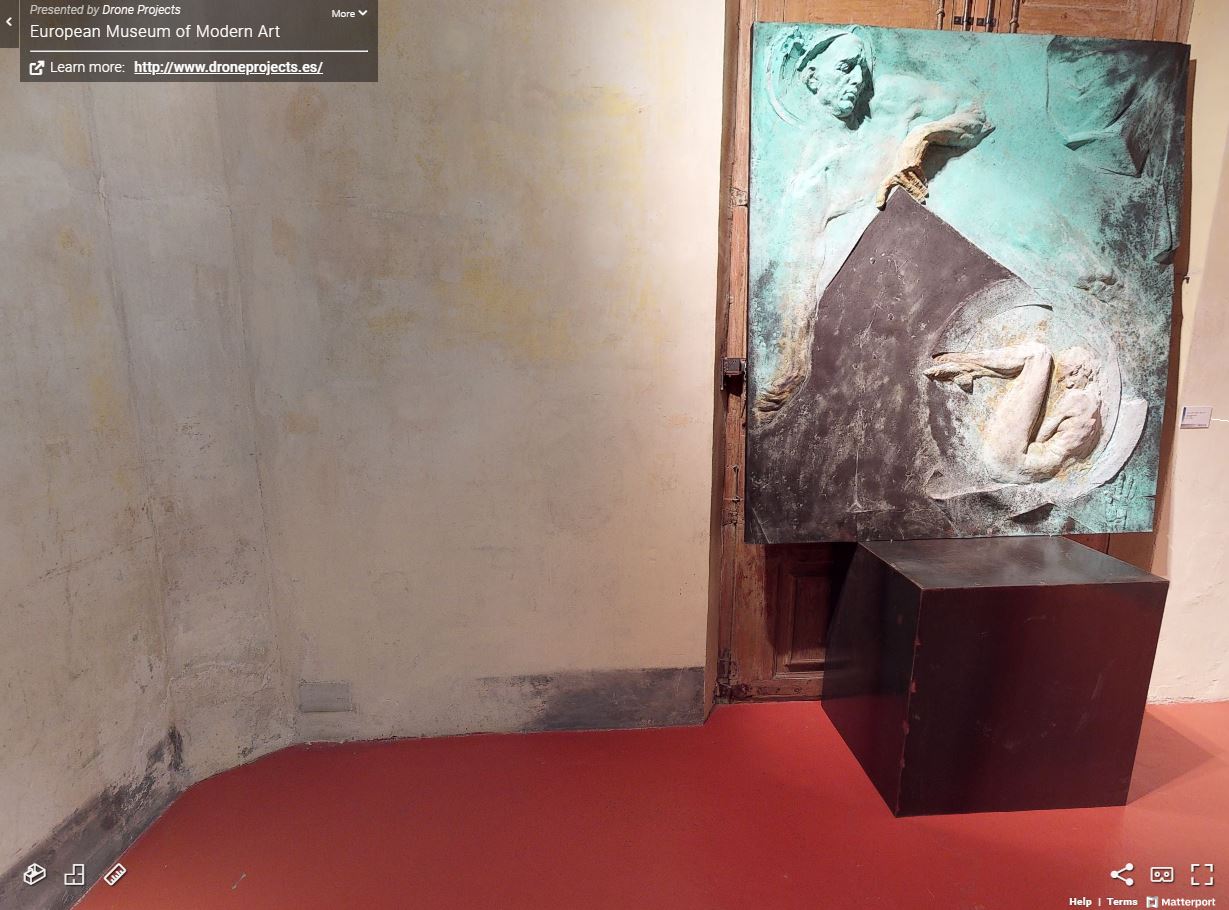} }}%
    \caption{Moving towards the left statue  does not take into account the gaze direction, resulting in the statue not being within the field of view.}%
    \label{fig:no-pov-adj}%
\end{figure}

% discrepancy - no POV adjustments vs expected POV
Another issue participants had problems with was the direction of gaze directly after commencing the movement. Namely, when the participants moved to a specific SNP near an exhibit, they expected to be placed in front of the exhibit with their direction of gaze adjusted \emph{toward} the exhibit. However, as a general rule, the system does not contextually adjust the participants' gaze direction. After transitioning participants to the desired SNP, the system simply assumed certain default gaze orientation that often does not take the exhibit into account. Consequently, the user may see the exhibit from an awkward or unintended angle, or even may not see the exhibit at all. An example for this can be seen in Figure \ref{fig:no-pov-adj}. The inability of the system to support contextual positioning and orientation created a loss of spatial awareness and disorientation for many participants. 

\begin{figure}[!htb]
    \centering
    \subfloat[\centering Regular field of view - no zoom applied, enables the user to pay attention to the SNPs and move closer to the exhibits ]{{\includegraphics[width=0.35\linewidth]{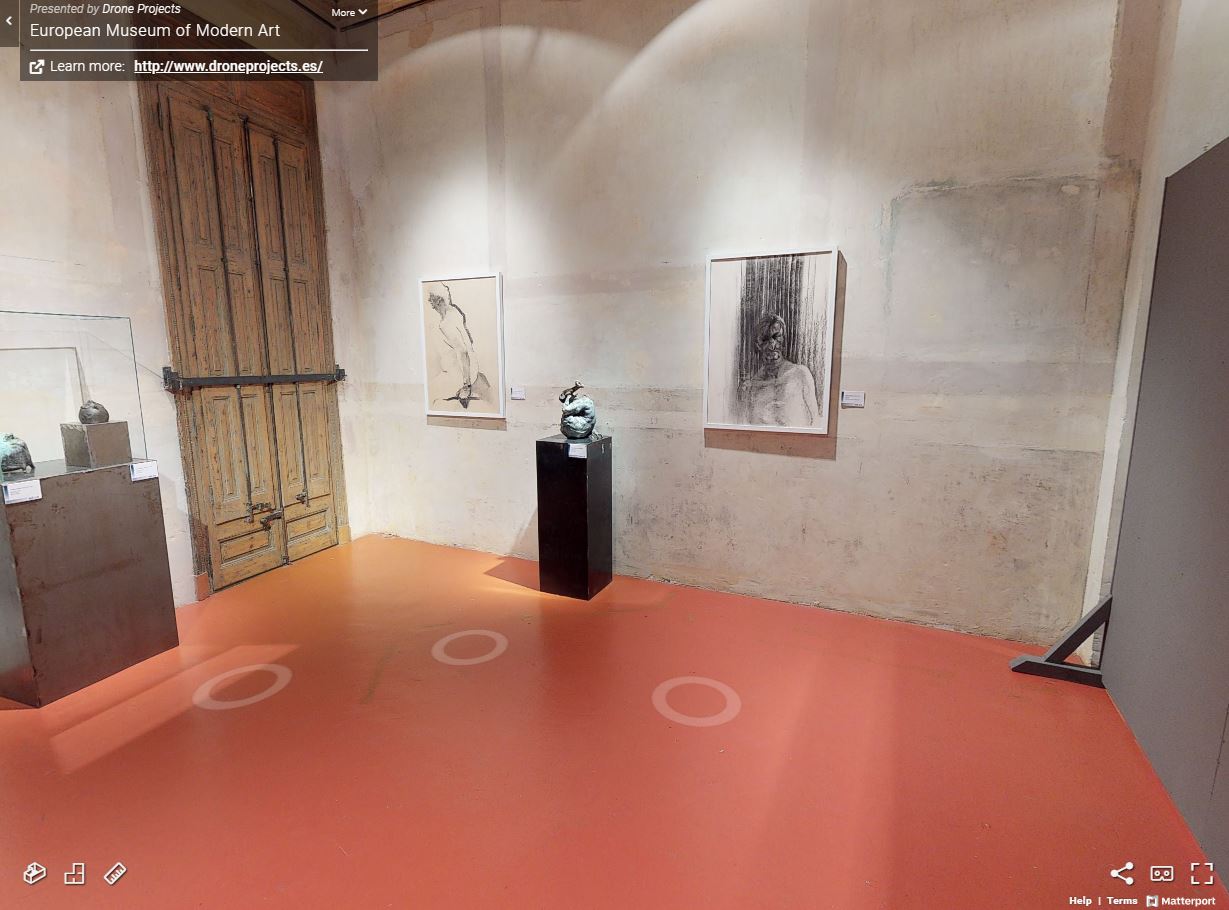} }}%
    \qquad
    \subfloat[\centering Narrow field of view - Excessive zoom applied to review the exhibit instead of moving closer ]{{\includegraphics[width=0.35\linewidth]{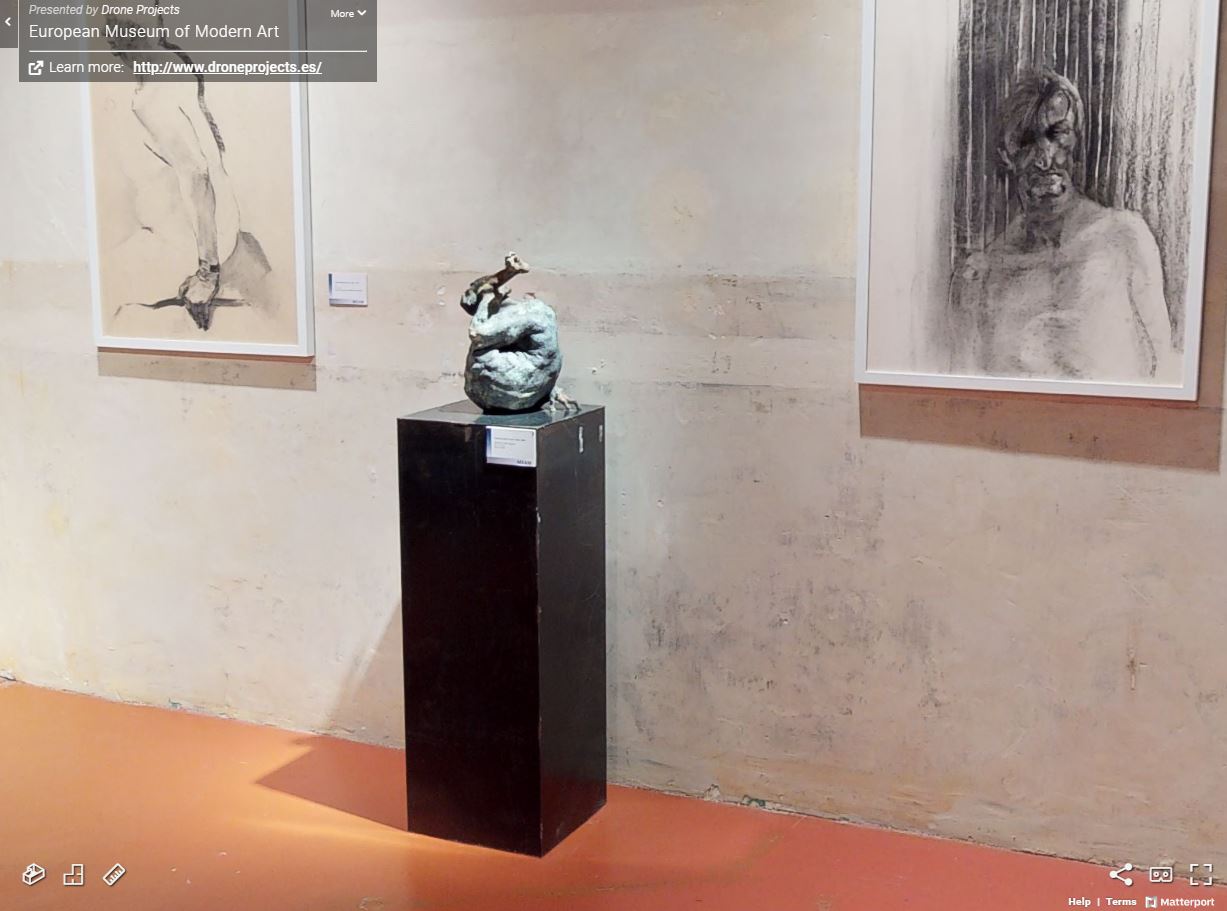} }}%
    \caption{Users tend to zoom instead of moving closer to exhibits, often resulting in staying zoomed-in when navigating, creating a narrow field of view}%
    \label{fig:tooMuchZoom}%
\end{figure}

%  Similarly = as in odd behavior. 
% stuck at zoom in/out
All 14 participants were observed to often use the zoom functionality to closely inspect exhibits. This often occurred when participants could not approach these exhibits because there were no available SNPs near them. Interestingly, many participants also appropriated the zoom functionality as a form of navigation mechanism that allowed them to "move" close to the exhibit, while simultaneously retaining their current static position. This is exemplified in the following statement by one of the participants: \textit{“I use the scroll wheel and I can approach and get close to it by zooming. It makes sense as scrolling forward moves me forward too”} (P5, 66).
However, similar to when looking down to find available SNPs, quite often, users had trouble or forgot to switch back when using the zoom function. As a result, such participants remained zoomed-in and continued the tour being partly or fully zoomed-in (observed with 8 of the 14 participants). What is especially interesting, is that these participants often did not realize that they are zoomed-in, as the system gave no indication of this fact. This created problems with effective navigation, since the users remained in a zoomed-in, limited field-of-view that made it more difficult to overview the museum environment, find available SNPs, and experience the space in a natural view (see Figure \ref{fig:tooMuchZoom}). 

% moving back / undo
Finally, we observed that moving backwards was  important to most of the participants. Participants often tried to get back to the point where they were standing before. There were several reasons for undoing the latest move, including moving ahead too much, losing spatial awareness, losing the point of view the user had on a specific exhibit, confusion in the navigation, and wanting to review an exhibit again. However, the conventional implementation of navigation in VTs does not support directly retracing one's previous steps or route. The participants tried to work around this limitation by manually turning back and searching for the previous SNPs. However, even when becoming proficient in point-based navigation, this is a difficult task. The participants did not always clearly understand where they came from and how to find their way back. Most digital systems require some sort of undo functionality which allows the user to reach the  previous state \citep{washizaki2002dynamic,abowd1992giving}. We saw that VTs are no exception, with the participants often attempting (and failing) to retract their steps.

\subsubsection{Limited Freedom of Navigation}
\hfill\\   
Even after participants understood \emph{how} to navigate, they often felt limited with respect to \emph{where} they could navigate and expressed their desire to be able to reach more areas and move more freely. Currently, existing systems mostly limit the number of SNPs around the exhibits, often leaving only one SNP per exhibit. As a result, in our observations, the participants were not able to examine the exhibits as they wished, choosing the appropriate angle and direction of view. The lack of SNPs around exhibits prevented users from appreciating and reviewing the exhibits from their own preferred position, as they would view them in a physical visit. 

\begin{figure}[!htb]
    \centering
    \subfloat[\centering Only one SNP is available near the exhibit, so the user cannot see the statue from all angles as in a museum ]{{\includegraphics[width=0.35\linewidth]{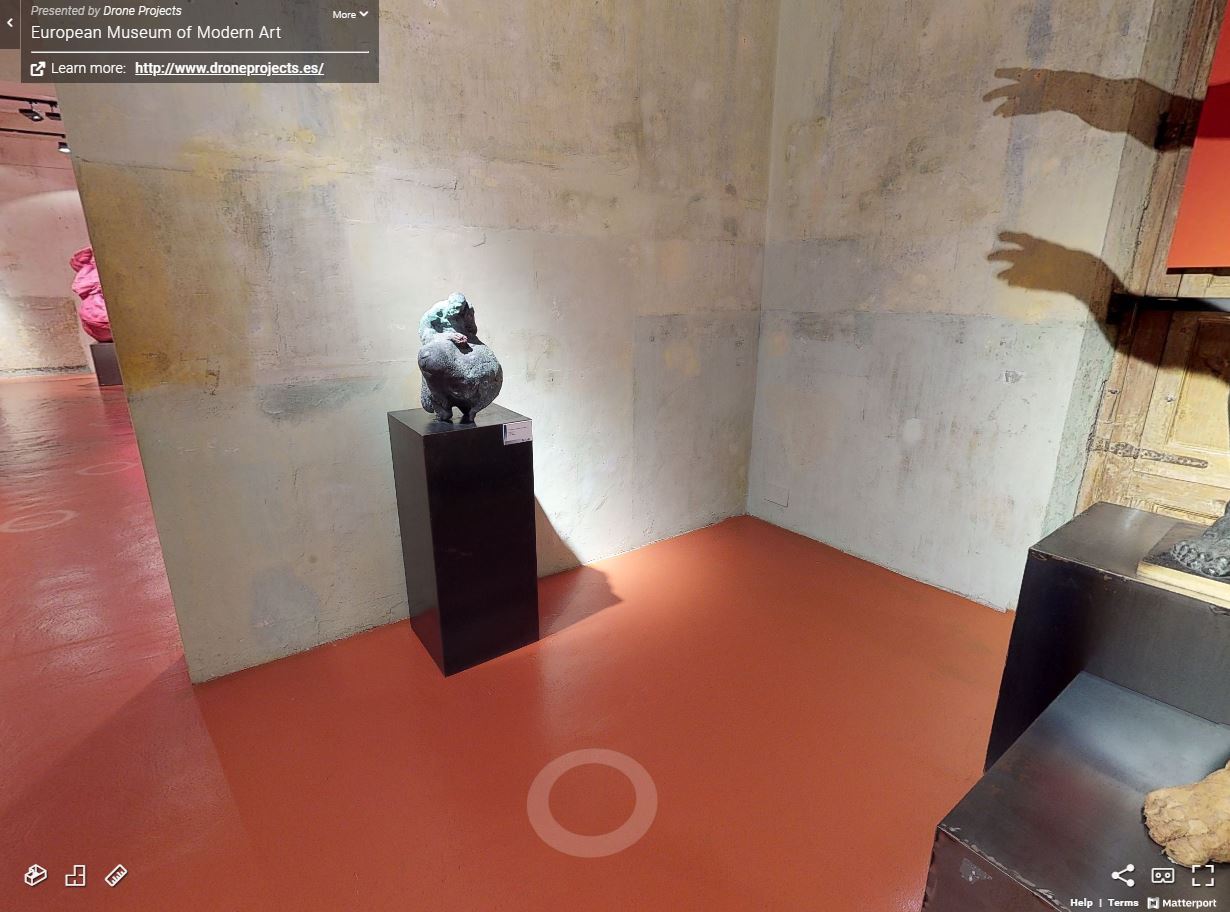} }}%
    \qquad
    \subfloat[\centering With several SNPs around the exhibit, the visitor can view it from  different angles. ]{{\includegraphics[width=0.35\linewidth]{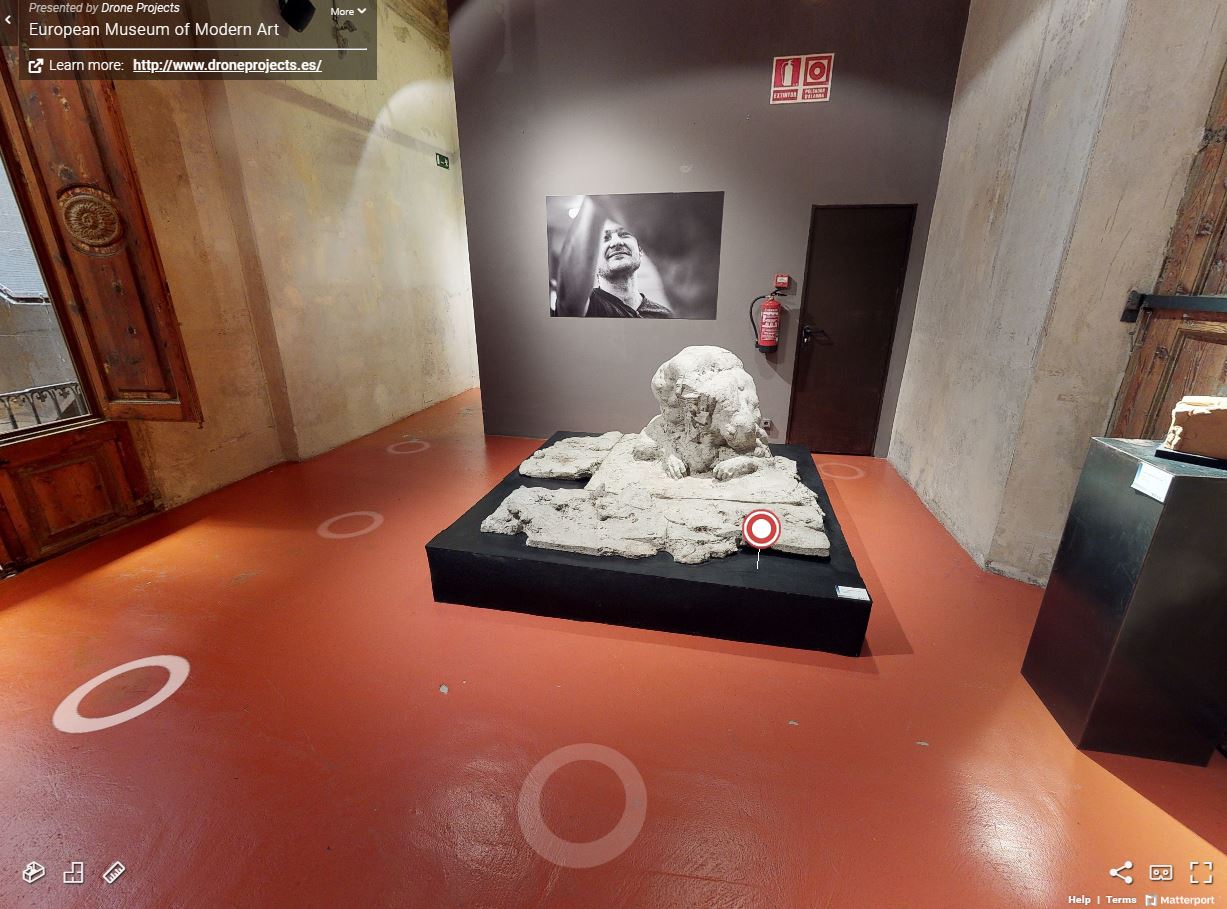} }}%
    \caption{more SNPs around the exhibit enables 360-degree view of the exhibit as in a real environment}%
    \label{fig:aroundExhibit}%
\end{figure}

P6 said that \textit{"I want the ability to move around this statue, however I am limited to moving to these points and it moves me away. I want a grid that will enable me to stand wherever I want, It'd be optimal. I cannot navigate entirely free, there are missing navigation points around this exhibit. [It is clear that] the digital medium is not a substitute."}. Rigid limitations on the available number of SNPs often created navigation problems as well since users tried to move around an exhibit, even if there were no SNPs to move to. Figure \ref{fig:aroundExhibit} shows an exhibit for which a user cannot look at from different angles (left) and one that the user can walk around (right).

While some exhibits did not have enough SNPs set around them, other exhibits did not have any SNPs near them at all, and thus were completely inaccessible to the users. Twelve (out of fourteen) participants complained about not being able to view certain exhibits. As one of the participants said:  \textit{"I feel that I cannot move freely. There are navigation points just for about half of the exhibits, and I want to move closer to them and I can't, it's annoying, what is the point if I cannot access half of the exhibits?"} (P3, 32 y.o.). 

In these cases, the participants did not have any choice beside using the zoom functionality to somehow look at the exhibit in more detail. However, as mentioned above, zooming-in on the exhibits often created spatial disorientation. Further, as the participants zoomed-in closer, the resolution of the image decreased, making the exhibits muddied and their details indiscernible. For example, P4 (33 y.o.) expressed that \textit{"I tried to review the exhibit from a convenient angle but it's not that easy to accomplish, I ended up zooming at the exhibit from far and it's not good enough in order to see it properly"}. 
 
\begin{figure}[h]
      \centering
      \includegraphics[width=0.35\linewidth]{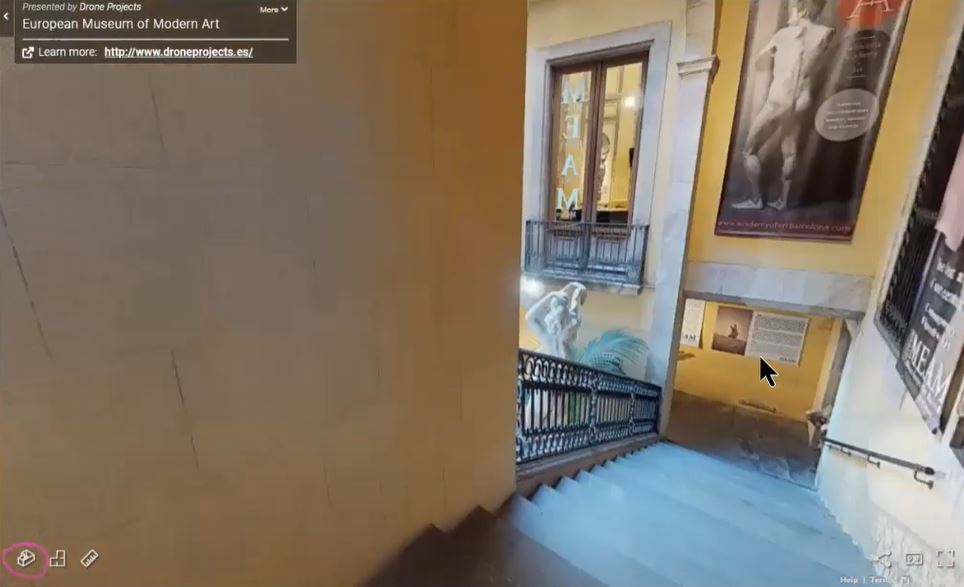}
      \caption{P3 was mislead to think that she can go down the stairs due an SNP near the stairs. however, when she pressed the stairs it moved her back to the room.}
      %\Description{Digital cues should not be provided to view from a distance}
          \label{fig:stairs}%
    \end{figure}

Finally, many participants (12/14) experienced difficulties in understanding the boundaries of the space that was modeled and included in the virtual museum tour. For example, in one of the tours, the museum environment included stairs leading to another floor. However, the second floor was not a part of the VT, and thus, was inaccessible to the virtual visitors (see Figure \ref{fig:stairs}), and the system did not provide any indication of the inaccessibility of the second floor. Consequently, the participants made several attempts to go down these stairs only to realize that this is not possible, contributing to further disorientation and frustration. As one of them explained: \textit{"clicking on the stairs and not being able to move there was frustrating, because I did not know that I should click on the navigation points only"} (P3, 32 y.o.). Similar situations emerged around closed doors and inaccessible entrances to other spaces.

%%%%%% add further
\subsubsection{Limited spatial awareness}
\hfill\\  
The participants often had difficulties understanding their location, orientation, and movement paths in the museum. This was evident from both participants' comments and from their behavior. We observed that many participants went back and forth in the same rooms in an attempt to understand where they are and where can they go. Still, most participants ultimately failed to form a coherent understanding of the space. For example, P6 asked \textit{"Are there more rooms I can go to? It is hard to me to understand how the museum is built"} (P6, 66 y.o.). 

It was clear to us that an additional navigational aid is required to support the participants' spatial reasoning and wayfinding. In the Matterport tour there was an overview map representing the floor plan of the VT's environment. However, it was hidden in the interface and participants were not aware of it. When we told them how to use it, the majority of the participants found this feature very useful for understanding the layout of the museum, and used it to orient themselves in space. Moreover, the map  also helped users to understand what rooms and spaces were available in the VT. However, we note that even with the map, the users still had trouble clearly understanding what rooms and exhibits they had previously visited, and what rooms and exhibits are still there to experience. 
As P4 (33 y.o.) has stated \textit{"I want to get an indication whether I was already in the room or not"}.

%Finally, tours often incorporate transitions between SNPs that were impossible or unnatural in fully physical spaces, which negatively affected the users' spatial awareness and orientation. For example, it is possible to go over statues from one SNP to another. In our observations, we saw how the participants were able to move to a certain SNP in another room by moving though walls. As P4 (33 y.o., male) reported "I've just went though a wall! it felt really weird". P3 (32 y.o. , female) complained that she got confused \textit{"I clicked on the statue and was teleported through the wall. I don't know what happened, it's not clear"}.

%%%%%%%%%%%%%%%%%%%%%%%%%%%%%%%%%%%%%%%%%%%%%%%%%%%%%%%%%%%%%%%%%%%%%%%%%%%%%%%%%
%%%%%%%%%%%%%%%%%%%%%%%%%%%%%%%%%%%%%%%%%%%%%%%%%%%%%%%%%%%%%%%%%%%%%%%%%%%%%%%%%
%%%%%%%%%%%%%%%%%%%%%% Information Presentation %%%%%%%%%%%%%%%%%%%%%%%%%%%%%%%%%%

\subsection{Information Presentation}
Beyond support for navigation and wayfinding, we observed that many of the participants' actions were centered around obtaining relevant contextual information about the exhibits that they were interested in. In this section we describe the participants behaviors and problems that have emerged when dealing with issues related to presentation of various types of information.

\subsubsection{Utilization of real-world information cues}
\begin{figure}[h]
      \centering
      \includegraphics[width=0.35\linewidth]{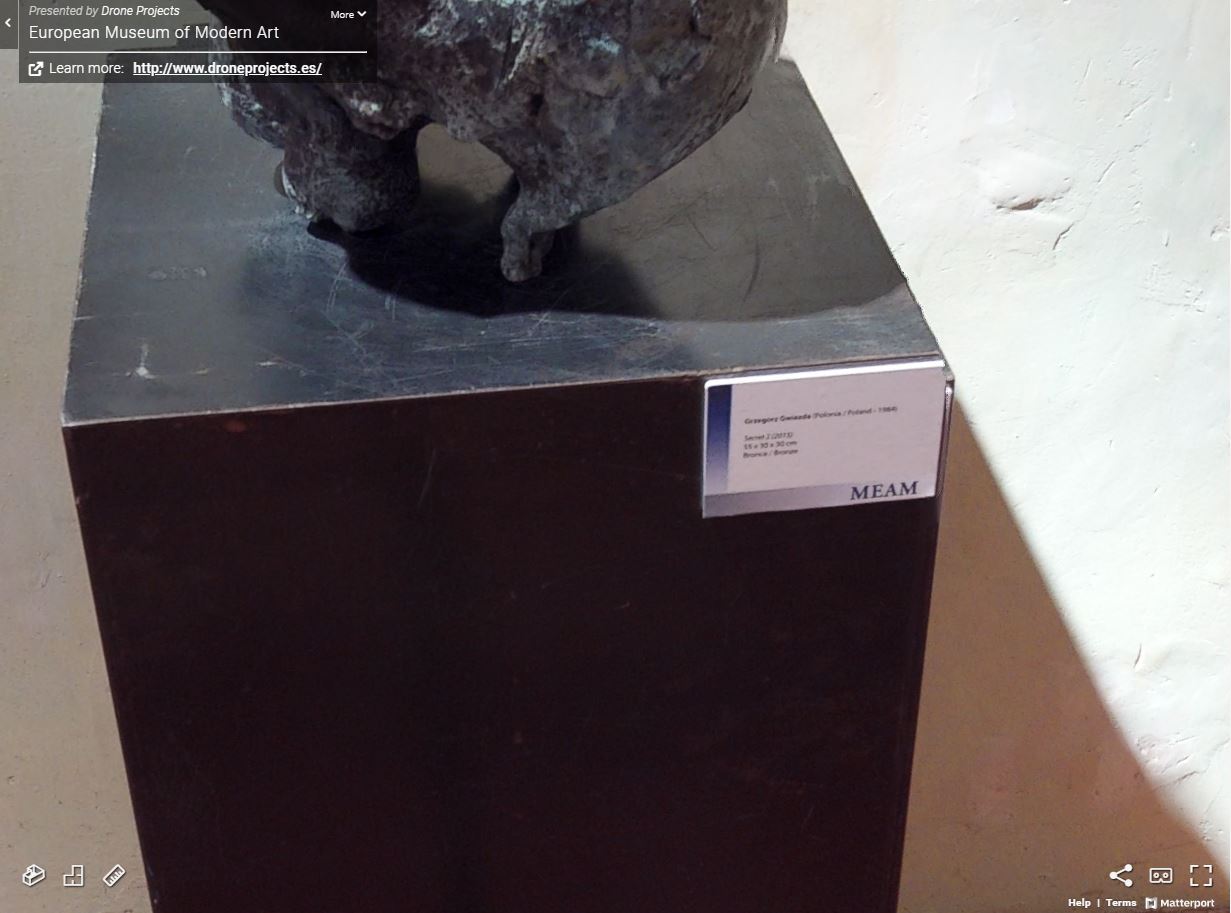}
      \caption{The participant attempted to zoom in to read the label, however even at maximum zoom ratio, the label is unreadable}
      %\Description{zoom to diegetic cue}
       \label{fig:zoom-to-diegetic-cue}%
    \end{figure}

\hfill\\
As with the case of navigation, we found a significant disparity between the expectations of the users and their actual experiences. We observed that often, our participants tried to read the actual physical labels of the exhibits to get information about them (this was observed with all 14 participants). These labels are typically scanned along with the rest of the physical environment when creating a VT. However, they are not intended by the designers as a significant information source for the VT visitors. This is evident from the low resolution of such labels that in most cases makes the text of such labels illegible. In addition, most labels in VTs are not directly approachable, as there are often no SNPs near these labels. Instead, VTs usually incorporate information on some of the exhibits in the form of an additional UI overlay window that the users initiate by pressing an icon positioned either near or on top of the exhibit (For example, see the red icon on the exhibit in Figure \ref{fig:aroundExhibit}(b)). 
Surprisingly, we discovered that even when an exhibit had a UI overlay with the relevant information, the users often attempted to read the “real” labels instead. However, as the labels were often too small or too far away from the SNPs, participants were not able to read them. Furthermore, the participants sometimes tried to zoom-in on such labels, further decreasing the legibility of the information on such labels (See Figure \ref{fig:zoom-to-diegetic-cue}). The ineffectiveness of the scanned labels caused frustration to the participants who realized that there was information there, but could not access it. As one of the participants stated: \textit{"I'm trying to review the exhibits in the showcase with their information beside them and it's impossible to read it. It's really a shame that I cannot get the information."}.(P4, 33 y.o.).
%P3 (32 y.o., female) complained that \textit{"I want to read the text but it's too small, and it seems to be impossible. I'd want to read the text beside the exhibit"}. 

A related behavior was observed when the participants attempted to use the physical directional and orientation signs for additional information about the place. For example, the participants looked for the names of the rooms, text on the walls, the exit and enter signs and other environmental elements. However, this was often difficult, as the museum signs were not modelled according to the VT's SNP locations. 
%In short, the participants, often unsuccessfully, looked for and expected to be able to use the various embedded cues that were part of the scanned environment for their informational needs. 

\subsubsection{Use of digital information} 
\begin{figure}[h]
      \centering
      \includegraphics[width=0.35\linewidth]{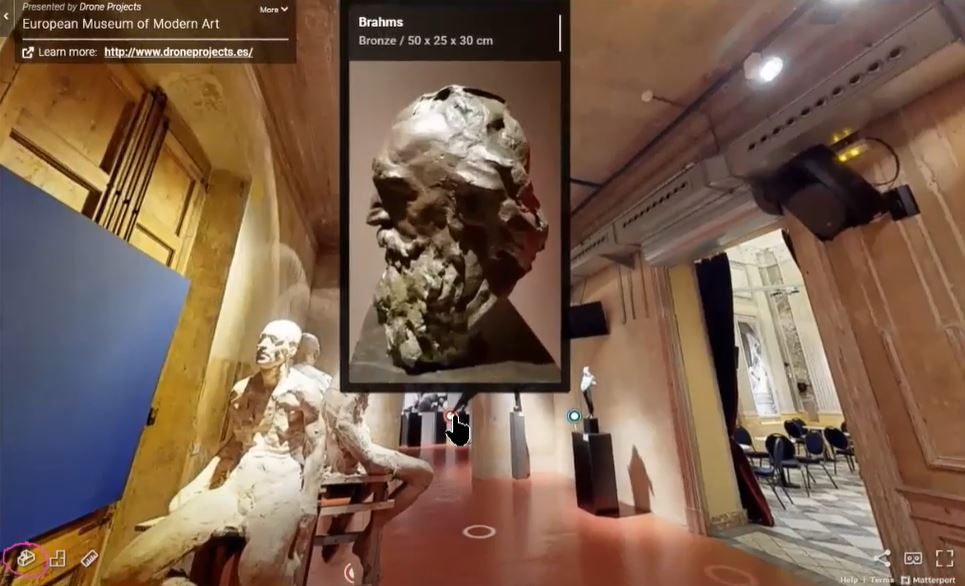}
      \caption{A participant viewing information on an exhibit which is located at the far end of the room. }
     % \Description{Digital cues should not be provided to view from a distance}
      \label{fig:digitalCue}
    \end{figure}

\hfill\\
As mentioned above, in both VTs, clickable icons were overlayed on top or beside  some of the exhibits to provide textual information on these exhibits when clicked. These digital cues could be seen throughout the room and could be pressed on from any location. We observed that often, participants pressed a cue to receive information about the exhibit, even when they were far away from that exhibit (see Figure \ref{fig:digitalCue}).  
In cluttered spaces, with many available digital cues, this created another problem. Since participants could see all digital cues in the room (even those that are obscured by other items), they were confused as to which cue belongs to which exhibit, especially when navigating between items and changing gaze directions.

The opposite phenomena also occurred and many participants complained that there is no information available on many of the exhibits. When we specifically asked, 13 out of the 14 participants said there was not enough information in the VTs. 
%P6 (69 y.o., male) expressed his dissatisfaction : \textit{"Some exhibits have information points and some do not. When I want to get closer and read what is written on the note -  I can not read" }.  
P8 (65 y.o.) expressed her frustration when she tried to access a statue and get information about it: \textit{ "I had expected that when I clicked on the statue, I would see a picture of it and be able to read about it - but that did not happen. [...] It is very interesting to me and I wish that I could learn more about it, but I am unable to do so and it frustrates me considerably."} It seems that because of the digital platform, participants expected to receive information on each and every exhibit and were disappointed when that was not the case.

%" i can tour the museum but i cannot understand what i see"
\subsubsection{Museum and room-level information} 
\hfill\\
Another requirement emerged around the need for general overview information about the museum content, and important rooms and exhibits available for the users, particularly at the start of the tour. Six of the 14 participants stated that they wanted to see a list of all the available rooms and exhibits in the museum tour at the start of the tour to be able to plan their visit more thoroughly. In general, such features are unavailable in typical VT experiences. Several participants noted that they also wished to have a list of \emph{important} exhibits or attractions to be available for reference. For example, P3 indicated that \textit{"I'd be happy to see a map of all the rooms, but it would not be enough. I want to receive a map of the exhibits, the exhibits in each room, their names, and [other stuff]"} (P3, 32 y.o.). 

\subsubsection{Multimedia Support}
Traditionally, textual information has been used as a go-to  medium in VTs to provide available information on specific exhibits. 
However, we note the general desire of many participants to implement additional forms of information. For one, 3D models of exhibits were mentioned as a possible way to provide information, with the ability to manipulate these models to examine exhibits up close. 3D models are particularly attractive because the identified issues with navigation around exhibits had led to the inaccessibility of many exhibits for close familiarization. Audio and video content was also a popular choice of the participants, described as a media that could positively contribute to the user experience. Multimedia can support a variety of information presentation approaches, from directly enriching informational labels, to complementary audio and video guides that ground the users' experience according to a specific narrative. P4 (33 y.o.) explicated: \textit{"I prefer that the exhibits would have associated audio guides, [...] I could choose to listen about the exhibit or read about it"}.

\subsection{Proactiveness}
The need for  \textit{interaction proactiveness} had emerged in our study as the expectation of the participants of the system in predicting their desire to engage with certain exhibits and proposing appropriate interactions. Some users suggested that the system would automatically trigger the appearance of the information card about the exhibit when the users look at this exhibit for a certain amount of time. For example, when one user went to look at an item that was particularly interesting to him, he actually waited several seconds, expecting the text about the exhibit to appear. He was quite disappointed when no text appeared: \textit{"It really frustrates me that I look at the exhibit and I am not provided with needed information about it"} (P12, 33 y.o.). Still, proactiveness should be implemented with care, as it takes control away from the users, and is not always the best approach to presenting museum digital content \citep{lanir2011examining}.

%%6

\section{Discussion}
Our findings provide a thorough look at users' behaviors and experiences when exploring online museum VTs. As with real museums, a positive user experience is vital for VTs success, as they are ultimately an experiential product aimed at both educating and entertaining the visitors. However, while a large body of knowledge exists around designing interactions in physical museum spaces, VTs so far attracted considerably less attention from the HCI and cultural heritage communities. 

Our findings indicate that there are many fundamental design issues that prevent the users from having a satisfying experience with VTs. Many of the participants' struggles had to do with general navigation and orientation issues. Users had difficulties figuring out where they are and how to find their way around the environment. Many of these navigation struggles had to do with the discrepancy between the users' mental model of how they are supposed to navigate in the VT and the actual system affordances. Our participants intuitively assumed that they can click on any point in the environment, expecting to advance to the exact point they have clicked on. This assumption was often not realized, creating general disorientation and confusion. Furthermore, the participants naturally tried to organize their navigation around the usage of the exhibits as landmark anchors that define and inform both their movement and orientation. However, the SNP implementation, which dictates the movement of visitors did not always adhere to the exhibits' locations and placements.

\par The uncovered problems originate from the idiosyncratic way of VTs creation. Because the designers need to "reconstruct" the real environment rather than simply model it in 3D, they cannot build a continuous virtual model of the environment similar to how video game worlds are created. Rather, a VT is constructed as a series of 360-degree sphere images that capture the real environment, and the users can only move between the centers of the "spheres" as navigation points. This necessarily creates a discrete movement experience which some users described as resembling going through a "flip book" of images.

\par Indeed, continuous movement was previously mentioned as one of the key factors to successful wayfinding in virtual spaces \citep{Malekmakan2020-jm}. VTs try to address this problem by implementing smooth transition effects when moving from one point to another. However, as we discovered, this approach only provides a partial solution to the experience of disjointed movement. It seems that the users, seeing a photo-realistic 3D space of a VT, innately expect a smooth navigation experience ("free movement tour"), similar to that which can be experienced through other virtual environments. Obviously, full modelling of the space could create a better navigation experience \citep{lepouras2001building,kersten2017development}. However, currently, full modelling is very costly and unpractical for most spaces and institutions. It appears to us that, given limited resources, the best solution for a better movement experience in VTs is to implement more SNPs in the environment, carefully placing them in conjunction to the structure and features of the scanned physical space. Such approach will enable small incremental movements throughout a rich navigational scaffolding and may alleviate some of the described navigational issues. 

\par Another way in which navigation and wayfinding could be supported is an implementation of guided navigation methods. In the context of fully virtual environments, Elmqvist and colleagues had previously described guided navigation as one of the effective solutions to support user's spatial cognition \citep{Elmqvist2008-we}. Thus, implementing an external map, clear information signs (showing which locations are off limits, and where it is possible to go) and navigation history (i.e., showing where the user has already visited) can aid spatial orientation in VTs. Another way to direct and inform movement is by using motion constraints \citep{Elmqvist2008-we}. In our context, pre-determined paths can be set along the VT's route, and gaze and viewing direction can be directed (e.g., according to the guidelines from Wernert and colleagues \citep{Wernert1999-pr}). However, this approach limits the users' freedom, and better fits museums in which the layout has a clear direct path throughout the space.

\par Finally, the use of exhibits as landmarks for navigation is an effective way to support orientation and wayfinding in a real museum \citep{wecker2015go}. Landmark-based navigation was also shown as an effective way for orientation in virtual environments \citep{Vinson1999-vg}. Similarly, in our study, the participants used exhibits to make sense of the space and navigate relative to the prominent and salient artifacts that they appropriated as landmarks. Thus,  designers should pay extra attention to the role of exhibits as navigation landmarks. One possible idea is to augment the central exhibits with virtual elements, making them stand out from the rest of the environment. In addition, implementing virtual landmarks may be an effective solution for VTs.

\par One of the particularly surprising findings was the participants' appropriation of diegetic cues (i.e., cues that are part of the scanned environment). Participants often tried to read the actual exhibit labels, instead of the more noticeable virtual labels. Furthermore, participants often looked at and used the directional signs on the wall in the museum rooms. This is especially interesting considering that other, much more salient, virtual orientation cues were overlayed on top of the reconstructed environment. Previous research on video-game design described how designers use both diegetic and non-diegetic navigational aids to support spatial cognition and orientation in 3D environments \citep{Dillman2018-lw}. Apparently, in the context of VTs, the spatial behaviors of the users could be much more aligned with the diegetic elements. Speculatively, such discrepancy could be associated with the visual realism of the VTs. If the users refer to such environments as "real", they may be more inclined to use diegetically available cues.

\subsection{Design Recommendations}
Based on our empirical findings, we present several concrete design recommendations, aimed specifically at designers and creators of VTs, to help create VTs that support users' spatial cognition and orientation, enhance users' interactions with the VTs', and result in a positive and memorable user experience.

%%%%%%%%%%%% 1 - Navigation Guidelines: 
    % more points : 360 review, continuous movement
\textbf{\textit{Plan and implement SNPs' placements based on the users' journey in the actual museum tour.}} To empower virtual visitors to explore the space and feel that they can move in a natural way, SNPs should be placed in all (seen) physical museum paths based on the identification of the users' routes and behaviors in the physical museum. This necessitates a close collaboration with museum curators and personnel, as well as knowledge of cultural heritage research, in order to incorporate their knowledge in the VT design process. 

\textbf{\textit{Distance between SNPs should be minimal.}} While the amount of SNPs might depend on the available resources, in general, the more SNPs the better. We observed that when moving between distant SNPs, and especially when moving between SNPs that are not aligned properly, participants often lost orientation due to the large jump that transfers them across large swaths of the environment. To be able to move gradually and fluidly in the space, the distance between SNPs should be minimal.

% better navigation signs
\textbf{\textit{Add signs that mark inaccessible places that are not mapped/included in the virtual tour.}} Not all museum areas can be mapped for the VT. However, the user should be informed if certain sections of the space are off limits or are not included in the tour. This can be done by placing clear virtual signs showing that these places are off limits. In addition we  recommend that SNPs should not be placed next to stairs, lifts and other locations that lead to off-limit areas, as this may mislead users into believing that they can be used to move ahead.
    
\textbf{\textit{Implement enough SNPs directly related to the exhibits and situated around them.}} Especially for the important exhibits, and those that can be viewed from different angles (e.g., statues), it is important to put several SNPs around them. This will enable users to view the exhibit from different angles as in a real museum. 

% make them move out of interest
%\item  Encourage Movement.  In terms of visual navigation guidance, VT navigation needs to be supportive, it should stimulate the visitor to proceed to the next accessible SNPs near the next exhibits, thus showing possible movements in an accessible and motivating way. Additionally, VTs should predict user's next moves, based on their preferences and past movement patterns and interests.
    
% constrain possible navigational mistakes
\textbf{\textit{Constrain the options for exhibit-based navigation}}. In both tours we examined in the current study (as well as most of the other VTs we reviewed earlier), the user can press any point on the screen in order to navigate to that direction. However, we found that this option is often confusing and unclear to the users. Thus, we recommend the system to only enable semantic exhibit-based navigation. That is, the system should not enable navigation when pressing random points (e.g., pressing on the wall), or exhibits that have no SNPs near them. The system should support exhibit-based navigation only for exhibits that are near SNPs.  

\textbf{\textit{Help users navigate using the correct gaze direction}}. Our findings indicate that often users remain during long periods of time in an inappropriate gaze direction, or excessively zoomed-in on the environment. To help users avoid this behavior, several steps can be taken. First, it is possible to implement a "realign" button, similarly to the functionality existing in many virtual navigation systems, where pressing a button restores one's default view. This can help users to realign their gaze direction and zoom levels at any time. Second, an icon or other visual elements indicating the current zoom level and the gaze orientation can be added to enhance the user's awareness. Finally, proactively realigning the user's gaze to the regular position can also be considered if the system detects the user is navigating for a long time in a less-than-optimal gaze orientation. %(e.g., the user navigates while looking at the floor all the time, or the user is navigating while being zoomed-in).   

\textbf{\textit{Support users' awareness of previous actions.}} Visualize and highlight previous navigation paths of the user on the map, or even within the environment itself, and highlight the spaces and rooms that were already visited. 
    
%\item Physical spatial anchors help users locate themselves in the space and indicate where they are. Therefore, consider placing and highlighting a physical or digital spatial anchor that makes it easier for users to locate themselves within the space. 

%%%%%%%%%%%%%%%%%%%%%%%%%%%%%%%%%%%%
    
%%%%%%%%%%%% 2 - Information Presentation Guidelines: 

    % information to all exhibits

\textbf{\textit{Design informative content on all main exhibits.}} While not directly related to the interaction design, most VTs that we reviewed, as well as those we used in the study, did not have enough information on their exhibits. This may have resulted in users trying (often unsuccessfully) to read the physical museum labels in order to receive some information on the exhibits. Thus, it is worth harnessing the inherent power of the digital medium, and create engaging, interactive textual and multimedia content for the museum exhibits as part of the VT. Such content can enrich users' interaction and learning without interfering with the museum's physical setting. 
    
\textbf{\textit{Avoid cluttering the space with redundant digital cues.}} While showing digital information for every exhibit is important, it can also create visual clutter that may affect the navigation and information consumption experience, especially in areas with a large concentration of exhibits. 
Thus, the presentation of digital cues and informational content could be organized according to the proximity of the users to the informational element. 
only those digital cues that are relevant to the user's current position should be shown. In addition to reducing the cluttering of the presentation, proximity-based presentation will aid users' spatial cognition. For example, if all the cues are reachable from any given place, when a user presses a far away cue (e.g., on the other side of the room), either by purpose or by mistake, it may break the spatial metaphor of the tour - that the virtual navigation follows the physical world rules. 
%Thus, we recommend that only those digital cues that are relevant to the user's current position should be shown.

%JL - need to see where this should be incorporated
%Therefore, to be technically accessible, all physical signs, notes, posters, and text must be visible and readable to VTists. Our results showed that virtual tourists preferred to read from diegetic cues such as plaques within the museum itself - rather than the special virtual UI prompts. However, if there are no physical cues or sufficient resolution to text reading , In order to serve the purposes of providing information to the user, interactive information cues must be virtually aligned with the actual physical description label, because this is the first and most natural location the users moves to in the intent of reading.

\subsection{Limitations}
Our work has several limitations. First, in order to scope our research, we focus on VTs using a Web interface to access the tour. However, there are other possible options to interact with VTs. Many users use their mobile devices as a portal to view the tour. Using a mobile device, such as a mobile phone or tablet, has very different affordances. On the one hand, the screen is smaller, and thus, it is more difficult to view the entire space. On the other hand, the interaction modality is different and uses touch to control and interact with the tour. In addition, natural movements (i.e., moving the phone from side to side) can be used for viewing the space. Examining the usability and experience of mobile phone users interacting with VTs remains future work. 

Virtual reality (VR) is another promising way to experience VTs, and various researchers examine how to best provide virtual experiences of cultural heritage sites to visitors via VR (e.g., \cite{kachach2020virtual,binti2020virtual}). Virtual reality can provide an immersive experience that can better simulate the real environment. However, VR is not yet accessible to most users, and may have issues such as motion sickness that prevent part of the users from adopting it. Still, the design and use of museum virtual tours in VR is an intriguing line of research.     

There are many types of VTs. Apart from museums, VTs can also be used for exploring real estate properties, campuses, exhibitions and other venues. In this work, we specifically focus on indoor museums as a typical use case for VTs. Museums usually include large rooms with many items of interest (i.e., exhibits). Other types of VTs, especially those who show disparate point of interests (e.g., real-estate VTs which usually show one image per room, with limited navigation), will most likely exhibit different behaviors and experiences.

%First, the behaviors and pain-points described in this study might not be exhaustive. 
In our study, we used two specific VTs with a sample size of 14 participants. We chose these two platforms as common and representative platforms for VTs today, and chose typical museum VTs. Still, the platforms and tours we used in the study  may affect the way participants interact with them. Looking at the participants of the study, a variety of demographics and ages were represented among the participants. However, although common in qualitative studies, 14 participants might not be a large enough sample to cover all possible user issues. 

%A variety of demographics and ages were represented among the participants. We analyzed the video sessions according to more than 60 indicators of activity levels, success or failure behaviours, and repeated usage patterns. Indicators were categorized in accordance with the earlier four-dimensional framework. 
%For each key indicator, we added a timestamp and documented their think-aloud in quotes. We focused in this study on the two main platforms for VTs use, that are commonly used and distributed. This study focused on the use of VTs on personal PCs, however future research will examine this in mobile applications and VR headsets \rs{and mobile devices such as smartphones and tablets}. 
Finally, previous experience with VTs might affect users as they might be more experienced with the navigation and the ways of interacting with VTs. However, in our study, we did not observe that past experience was a factor in user success. In fact, we saw the same behaviors in both experienced and non-experienced users. Furthermore, we did not observe much learning or change in behavior as the tour went on. Future work should examine whether training and more experience of users might affect the overall user experience of VTs.

\section{Conclusion}

Virtual tourism applications aim to bring  cultural heritage sites to the homes of the visitors, making them more accessible to larger audiences. 
%It is therefore time to explore and deeper examine the various possibilities of virtual tourism.
VTs of museums and cultural heritage sites can be a great way to experience virtual tourism. Museum VT visitors can view the museum collection and can learn about the artifacts and the exhibitions similar to the way they would in a real museum. Still, similar to the amount and effort put in designing a physical exhibition, a virtual exhibition needs to also be carefully thought of, and there is no doubt that the user experience of the museum VT is critical to its success. 

Our goal in this work was to take a look at the current status of the design of virtual tours in order to  understand how to better design future VTs. We did this by providing a descriptive framework of VTs based on the analysis of 40 VTs, as well as a user study that examined the usability issues and user pain points when using two typical VTs. 
Our results highlight users' navigation struggles which were often caused by a mismatch between their mental models of how the system should work and its actual capabilities. Furthermore, we show how users view virtual tours as realistic representations of museum experiences, and are therefore more likely to utilize the diegetic cues available. 
%One clear conclusion is that users' spatial behaviors during virtual tours need to be better matched with diegetic elements to improve user experience.

%We provide several actionable design recommendations for VTs based on our empirical findings, aimed to help designers and developers of VTs, in order to develop VTs that facilitate users' spatial cognition and orientation, enhance their interactions with the VTs, and result in a positive user experience.

In conclusion, virtual tours play a crucial role in enhancing public accessibility to museums and cultural heritage sites. Our research delved into the impact of VT design on visitors' overall user experience, uncovering valuable insights and providing actionable design recommendations. These findings hold great significance for researchers, designers, and developers working on VTs, as they contribute to the continuous improvement of virtual tourism user experiences.

\bibliographystyle{apacite}
\bibliography{interactapasample}

%\section{Appendices}
%\appendix

\end{document}